\documentclass[a4paper,12pt]{article}

\usepackage{iftex}
\ifPDFTeX
\usepackage[utf8]{inputenc}
\usepackage[T1]{fontenc}
\else
\usepackage{fontspec}
\fi

\usepackage{amssymb,amsfonts,bm,mathtools} 
\usepackage{geometry}
\usepackage{graphicx}
\usepackage{enumitem}
\usepackage{setspace}
\usepackage{booktabs}
\usepackage[hidelinks]{hyperref}
\usepackage{float}       
\usepackage{ulem}
\usepackage{subcaption}  
\usepackage{pgf}
\usepackage{placeins}
\usepackage{pdflscape}
\usepackage{adjustbox}
\usepackage{multirow}
\usepackage{caption}
\usepackage{fancyhdr}

\numberwithin{equation}{section}
\numberwithin{figure}{section}
\numberwithin{table}{section}
\newcommand{\underTilde}[1]{\ensuremath{\boldsymbol{#1}}}

\begin{document}
	

	
	
	\begin{center}
		{\Large\bfseries
			A Revisit to Point Estimation Through the Empirical Bayes Method: The Case of Binomial Distribution with Beta Prior and Extension to Poisson Distribution
		}
		
		\vspace{10pt}
		
		\textbf{Quoc-Bao Nguyen}$^{1,2,3,\dagger}$,
		\textbf{Nabendu Pal}$^{3,*}$,
		and \textbf{Dang Van Vinh}$^{1,2,\ddagger}$
		
		\vspace{8pt}
		
		{\footnotesize
			$^{1}$ Faculty of Applied Science, Ho Chi Minh City University of Technology (HCMUT),\\ 268 Ly Thuong Kiet Street, Dien Hong Ward, Ho Chi Minh City, Vietnam\\
			$^{2}$ Vietnam National University Ho Chi Minh City, Linh Xuan Ward, Ho Chi Minh City, Vietnam\\
			$^{3}$ Faculty of Mathematics and Statistics, Ton Duc Thang University, Ho Chi Minh City, Vietnam\\
		}
	\end{center}

	\vspace{1em}
	
	\begin{abstract}
		Between the classical (frequentist) approach, which is based solely on the data, and a fully Bayesian set-up where one assumes a prior distribution for the model parameters, lies the Empirical Bayes (EB) approach which appears to be a good compromise between the aforementioned two approaches. Even though many researchers have suggested various variants of the EB method, the standard practice is to derive the Bayes estimator under a family of suitable priors indexed by its own parameter(s), called the hyperparameter(s), and then replace the unknown hyperparameter(s) by their estimate(s) obtained from the marginal distribution of the data. But the fundamental question that is being raised here is: does the EB method really work to produce an improved estimator - the so-called Empirical Bayes Estimator (EBE)?
		
		In this work we are going to revisit the widely cited simple problem of estimating a Binomial parameter using the regular two-parameter Beta family of priors under the quadratic loss function, and prove that the Type-II maximum likelihood (ML-II) step does not work. If we further restrict our attention to one-parameter symmetric Beta family of priors then still the resultant EBE does not show any remarkable performance compared to the MLE details of which have been provided with extensive computations. The Binomial study has been extended to the Poisson model as well.
	\end{abstract}
	
	

	\vfill
	\hrule
	\vspace{0.2cm}
	
	\footnotesize
	
	\noindent
	$\dagger$ First author email:
	bao.nguyen2580892@hcmut.edu.vn;
	nguyenquocbao1@tdtu.edu.vn
	
	\noindent
	$*$ Corresponding author email:
	nabendu.pal@tdtu.edu.vn
	
	\noindent
	$\ddagger$ Third author email:
	dangvvinh@hcmut.edu.vn
	
	
	\normalsize
	\newpage

	\section{Empirical Bayes Estimation: Background}
	\subsection{Preliminaries}
	
	There is an endless debate about the selection of a suitable prior distribution within a Bayesian set-up. In this regard, the 
	Empirical Bayes method seems to be a good compromise between a fully Bayesian set-up and the 
	classical (frequentist) approach. In a nutshell, let us assume that the data, say $\underTilde{X}$, given the 
	parameter $\theta$, follows the model $f(\underTilde{x}|\theta)$, i.e., $\underTilde{X}|\theta \sim f(\underTilde{x}|\theta)$, 
	$\theta \in \Theta$, where $\Theta$ is the parameter space defined as $	\Theta = \{\, \theta \!\mid\! {\textstyle\int} f(\underTilde{x}|\theta) \, d\underTilde{x} = 1 \,\}$, where $\theta$ may be multi-dimensional, and hence $\Theta \subseteq \mathbb{R}^d$, for some suitable $d \ge 1$.
	
	We further assume a prior distribution for $\theta$ as $\pi(\theta)$, i.e. $\int_{\Theta} \pi(\theta)\, d\theta = 1,$
	where the prior $\pi(\theta)$ may depend on the hyperparameter $\eta$, and hence if that ever happens (which is going to be the case in this work), we are going to denote the prior as $\pi(\theta | \eta)$ . In a pure proper Bayesian set-up it is assumed 
	that $\eta$ is completely known, which makes it a bone of contention. By allowing $\eta$ to take a suitable value over its natural range, say $\mathcal{H}$, where
	$\mathcal{H}
	=
	\left\{\eta:
	\int_{\Theta}\pi(\theta\mid\eta)\,d\theta=1
	\right\},$
	one can bring some degree of flexibility in the prior selection, and hence to show the prior's dependence on $\eta$ over the space $\mathcal{H}$ we adopt the notation $\pi(\theta|\eta)$, $\eta \in \mathcal{H}$, instead of $\pi(\theta)$.
	
	In order to draw inferences on $\theta$, one uses the posterior distribution of $\theta$ after observing $\underTilde{X}=\underTilde{x}$. Let
	$m(\underTilde{x},\theta\mid\eta)
	=f(\underTilde{x}\mid\theta)\pi(\theta\mid\eta)$
	and
	$m_1(\underTilde{x}\mid\eta)
	=\int_\Theta m(\underTilde{x},\theta\mid\eta)\,d\theta$
	denote the joint and marginal distributions, respectively.
	The posterior distribution of $\theta$ given $\underTilde{X} = \underTilde{x}$ is 
	$
	\pi(\theta \!\mid\! \underTilde{x}, \eta) = m(\underTilde{x}, \theta \!\mid\! \eta)/{m_1(\underTilde{x} \!\mid\! \eta)}.
	$
	
	Under the squared error loss function, the optimal estimator which minimizes the Bayes risk is the posterior mean, i.e.,
	\begin{equation}
		\hat{\theta}_B(\underTilde{x} \!\mid\! \eta) 
		= \int_{\Theta} \theta\, \pi(\theta \!\mid\! \underTilde{x}, \eta)\, d\theta.
		\label{eq:bayes_estimator}
	\end{equation}
	
	In an Empirical Bayes set-up, one only assumes the functional form of $\pi(\theta\!\!\mid\!\!\eta)$, 
	but keeps the hyperparameter $\eta$ open which needs to be estimated from the data $\underTilde{X}$ 
	using the marginal distribution $m_1(\underTilde{x} \!\mid\! \eta)$. Typically, if $\hat{\eta}$ 
	is an estimator of $\eta$ based on the marginal distribution $m_1(\underTilde{x} \!\mid\! \eta)$, 
	then by plugging it in \eqref{eq:bayes_estimator} yields an Empirical Bayes estimator of $\theta$ as
	\begin{equation}
		\hat{\theta}_{EB}(\underTilde{x}) = 
		\hat{\theta}_B(\underTilde{x} \!\mid\! \hat{\eta}).
		\label{eq:eb_estimator}
	\end{equation}
	
	Most of the time one uses $\hat{\eta}_{ML-II}$, the ``Type-II maximum likelihood estimator'' (ML-II), 
	which maximizes $m_1(\underTilde{x} \!\mid\! \eta)$ with respect to $\eta$ (see page-99, Berger (1985)). However, conceptually 
	it is possible to use other types of estimators of $\eta$ as well, such as the method of moment(s) 
	estimator (MME) of $\eta$ in \eqref{eq:eb_estimator}.
	
	\subsection{A Brief Literature Review on Empirical Bayes Method}
	
	One of the earliest works on EB estimation is of
	Robbins (1956). Among other things this work considered the	Poisson distribution, and argued that the risk of the EBE would approach the Bayes risk as the number of observations increases. Martz and Lian (1974) considered a sequence
	of experiments where, in the $i$-th experiment $(1\le i\le m)$,
	$X_i$ successes
	were observed in $n_i$ trials. Further, $P_i$'s were assumed
	to be $iid$ realizations from an unknown prior, say $G(p)$.
	The goal was to use information from previous experiments
	to improve the estimation of the current parameter through
	a smooth EBE called the 'Smooth Incomplete Beta Estimator'
	(SIBE), even though this estimator showed a somewhat
	good risk performance when $m$ and $n_i$'s were large.
	
	Gutmann (1982) considered the same setup as that of
	Martz and Lian (1974) with the goal not to construct a
	new estimator, rather to determine when information from
	past observations should be used and when only the
	current data ought to be used. The author justified the proposed method using Efron and Morris's (1975) baseball data and 	the Portsmouth Naval Shipyard data.	Efron (2019) presented a methodological review of Bayes,
	Oracle Bayes (the ideal benchmark when the true prior $\pi$ is treated
	as known) and EB methods. The works considered the
	general framework with $X_i \mid \theta_i \sim p(x_i \mid \theta_i), $
	where $\theta_i$ is an
	unobserved parameter. It was assumed that $\theta_i$'s are
	generated from an unknown prior $\pi(\theta)$, and the data have
	marginal distribution
	$
	m(x)
	=
	\int p(x \mid \theta)\,\pi(\theta)\,d\theta.
	$
	The main contribution here is to clarify the dual nature of EBE.
	On the Bayesian side, the papers considered posterior
	inference for each parameter after the prior had been estimated. On the frequentist side, it was argued that using $m(x)$ could be advantageous when the marginal could be learned well from the data. Jana et al.\ (2025) considered independent Poisson data
	$
	X_i \mid \theta_i \sim \mathrm{Poisson}(\theta_i),
	1 \le i \le m, $
	where $\theta_i$ are supposed to be $iid$ from $\pi(\theta)$. The goal had been to estimate the prior from the data,
	then construct a Bayes estimator followed by evaluating its
	performance relative to Oracle Bayes (that knew the true prior).
	Chen and Lei (2025) took a different approach by approximating
	Stein's Unbiased Risk Estimator (SURE) for multiple Binomial parameters.
	In a multivariate Gaussian set-up, SURE is already a well-established
	tool, but not so easy in the Binomial setting. These authors proposed a class of shrinkage estimators combining the classical MLE, grand mean, and machine-learning predictor. 	
	A somewhat different approach was taken by Kang et al.\ (2026),
	where the authors considered estimation of functions of
	parameters in the nonparametric EB setting. The two
	main models studied were Poisson and Gaussian. For an excellent, yet brief, expository 
	discussion on the EB method see Carlin, B. P. and Louis, T. A. (2000), and other useful references therein.
	
	It appears that the EB method tends to work well, in terms of providing
	improved estimation over the classical ones, when the number of model
	parameters is more than a critical value (`dimension'), and this has
	provided justification for the existence of `Stein-type' shrinkage
	estimators (see Casella (1985) for detailed intuitive justifications).
	However, it should be kept in mind that EB justification was introduced
	later, while Stein's original uniformly improved shrinkage estimators
	appeared in the literature much earlier using a completely different
	argument. This argument also worked for simultaneous estimation of
	several Poisson parameters and Gamma scale parameters; see Ghosh and
	Parsian (1981) and Berger (1980), respectively. Interestingly, although
	EB arguments have been used to justify uniformly superior shrinkage
	estimators for the above distributions, the same intuitive argument
	does not appear to work for simultaneous estimation of several Binomial
	parameters, because uniformly superior shrinkage estimators do not exist
	in this case; see Johnson (1971).
	
	Coming back to the EB problem, doubts still persist about how to estimate
	the hyperparameter(s) from the marginal distribution of the data,
	i.e., \(m_1(x\mid\eta)\). Maritz and Lwin (1989) considered EB estimation
	for several common distributions, including the Binomial distribution
	(see p.~101), and suggested both the ML-II approach and the method of
	moments (MM) based on the marginal distribution of the data. (Analogously, the MM procedure applied to the marginal
	distribution may be referred to as ‘MM-II’.) However, we will see in the subsequent sections
	that neither method really works for the Binomial problem considered here with the Beta prior.
	
	Doss and Linero (2024) did study the implementation of the EB method
	through the maximization of the marginal likelihood function, i.e.,$
	\widehat{\eta}_{\mathrm{ML-II}}
	=
	\arg\max_{\eta} m_1(x\mid \eta).
	$
	They argued that outside some simple textbook examples,
	\(m_1(x\mid \eta)\) and its maximizer are analytically intractable.
	That is why they proposed an MCMC-based approach by introducing a fully
	Bayesian auxiliary model by placing a hierarchical prior on the
	hyperparameter \(\eta\), and then running an MCMC chain on \((\theta,\eta)\).
	However, in light of our dealing with the simple problem of binomial distribution with Beta
	family of priors, and the difficulty of obtaining a meaningful EBE as we have experienced, the
	utility of Doss and Linero’s (2024) MCMC approach becomes questionable. Many a time, the
	functioning of a computational algorithm, however noble it might be under ideal assumptions, turns 
	out to be a black-box, and it is not clear whether the ultimate outcome that we are getting is truly 
	the one desired for, or is it just some hodgepodge cooked up by the algorithms’ alchemy which 
	is quite different from the one that we were supposed to get.
	
	The algorithmic black-box mystery can be demonstrated by a completely different example. There has been a great interest lately to generalize the usual univariate Normal distribution by
	a more flexible Skew-Normal distribution (SND) which incorporates an extra skew parameter besides the location and scale parameters (see Azzalini (2013) who has done some pioneering work on SND). Multivariate generalizations of SND have been in use for sometime which is an active area of research. Interestingly, given a dataset, finding the MLEs of the three SND model parameters,
	especially the skew parameter, is a challenge even in the univariate case. (Estimation of the location as well as the scale parameters is directly related to that of the shape parameter.) When all the
	observations fall on one side of the location parameter (or, in the likelihood function setting, when the
	location parameter is confined to either on the left side of the smallest observation or on the right
	side of the largest observation), then the likelihood function gets maximized as the skew parameter
	approaches negative or positive infinity. In other words, MLEs of the SND model parameters do not
	exist because of the aforementioned regions with however small probabilities they may carry.
	Yet, in a multivariate setting, EM Algorithm has been proposed to find the 'MLE's of the model
	parameters (see  Arellano-Valle et al. (2018)) where the parameter MLEs have been obtained
	for a real-life dataset.
	
	In the following section 2, we explore the Binomial distribution with its conjugate Beta prior and examine whether ML-II provides finite estimates of the hyperparameters. We show that it does not in the one-observation setting considered here. We then study the related family of Beta posterior-mean estimators with fixed hyperparameters and identify choices that improve upon the MLE in maximum risk, though not uniformly in pointwise risk, under squared error loss.
	
	\section{Empirical Bayes Estimation of a Binomial Parameter}
	
	Let $X|\theta \sim \mathrm{Binomial}(n, \theta)$, and the conjugate prior $\pi(\theta)$ is the usual $\mathrm{Beta}(a,b)$ distribution. Hence,
	$	X\!\mid\!\theta \sim f(x\!\mid\!\theta)=C_x^n\,\theta^{x}(1-\theta)^{\,n-x},
	\quad x\in R_X = \{0,1,2,\ldots,n\}$, and
	$
	\theta \sim \pi(\theta \!\mid\! a,b) = \{1/B(a,b)\} \theta^{a-1}(1-\theta)^{b-1}, \theta \in \Theta = (0,1),\ a>0,\ b>0,
	$
	\noindent
	\noindent
	where $B(a,b)$ is the usual Beta function with the representation  $B(a,b) =\Gamma(a)\Gamma(b)/\Gamma(a+b)$. The joint distribution of $(X,\theta)$, denoted by $m(x,\theta|a,b)$, is
	$
	m(x,\theta|a,b) = C_x^n \{1/B(a,b)\}\, 
	\theta^{a+x-1}(1-\theta)^{b+n-x-1},
	$
	which yields the corresponding marginal distribution of $X$ as 
	\begin{equation}
		m_1(x|a,b) = C_x^n\,
		B(x+a,\, n-x+b)/B(a,b),
		a>0,\ b>0,\ x \in R_X =  \{0,1,2,\ldots,n\}.
		\label{eq:beta_binomial_marginal}
	\end{equation}

	\noindent The posterior distribution of $\theta$, after observing $X=x$, is
	$
	\pi(\theta|x) = \{1/B(x+a,\, n-x+b)\}\, 
	\theta^{x+a-1}(1-\theta)^{n-x+b-1}, 
	\theta \in \Theta,\ a>0,\ b>0.
	$
	The Bayes estimator, which minimizes the Bayes risk under the squared error loss function, 
	is the posterior mean, which in this case is
	$
	\hat{\theta}_B(x|a,b) = (x+a)/(n+a+b).
	\label{eq:beta_binomial_bayes_estimator}
	$	
	In an EB set-up, $a$ and $b$ are the hyperparameters and should be estimated from 
	the marginal distribution \eqref{eq:beta_binomial_marginal}, which yields 
	\begin{equation}
		\hat{\theta}_{EB}(x) = (x + \hat{a})/(n + \hat{a} + \hat{b}).
		\label{eq:beta_binomial_eb_estimator}
	\end{equation}
	For convenience we will write $\eta = (a,b)$ in subsequent sections.
	
	\subsection{The ML-II Method: Does it Work with General Beta Prior?}
	
	In the following, we investigate $\hat{\eta} = (\hat{a}, \hat{b})$ and the resulting estimator $\hat{\theta}_{EB}$ under the ML-II approach, and compare it with the standard estimator
	$\hat{\theta}_0 = x/n$
	which is not only the MLE of $\theta$, but also its unique 
	minimum variance unbiased estimator (UMVUE).
	
	First, we try to find $(\hat{a}, \hat{b})$, which maximizes $m_1^*(x|a,b)$ after dropping the multiplier $C_x^n$ 
	(which is free from $(a,b)$), where 
	\begin{equation}
		m_1^*(x|a,b) = \frac{B(x+a,\, n-x+b)}{B(a,b)} 
		= \frac{\Gamma(x+a)\, \Gamma(n-x+b)\, \Gamma(a+b)}
		{\Gamma(n+a+b)\, \Gamma(a)\, \Gamma(b)}
		\label{eq:reduced_marginal_likelihood}
	\end{equation}
	for the following two special cases.
		
	\subsubsection{The special case of $n$ = 1}
	
	For \(n=1\), equation \eqref{eq:reduced_marginal_likelihood} becomes
	$
	m_1^*(x|a,b)
	=
	\{\Gamma(x+a)\Gamma(1-x+b)\Gamma(a+b)\}/
	\{\Gamma(1+a+b)\Gamma(a)\Gamma(b)\} 
	$ for $x\in\{0,1\}$. If \(x=0\), then from \eqref{eq:reduced_marginal_likelihood},
	$
	m_1^*(0|a,b)
	=
	\left(1+a/b\right)^{-1},
	$
	which approaches its supremum as \(a/b \to 0\), that is, when \(a\to 0\) and/or \(b\to\infty\). When \(x=1\), from \eqref{eq:reduced_marginal_likelihood},
	$
	m_1^*(1|a,b)
	=
	(a/b)/(1+(a/b)).
	$
	Hence,
	$
	\sup_{a,b>0} m_1^*(1|a,b)=1,
	$
	and the supremum is approached as \(a/b\to\infty\), i.e., when \(a\to\infty\) and/or \(b\to 0\).
	
	These limiting values lie outside the finite hyperparameter space \(a,b>0\). Along the corresponding limiting paths, the value obtained from \eqref{eq:beta_binomial_eb_estimator} agrees with the MLE \(\hat{\theta}_0=x/n\). Thus, ML-II does not provide finite estimates of \(a,b\) for this Bernoulli case.
	
	\subsubsection{The special case of $n$ = 2}
	From \eqref{eq:reduced_marginal_likelihood},
	$
	m_1^*(x|a,b) = \{\Gamma(x+a)\Gamma(2-x+b)\Gamma(a+b)\}/
	\{\Gamma(2+a+b)\Gamma(a)\Gamma(b)\}.
	$
	At $x = 0$, \eqref{eq:reduced_marginal_likelihood} yields
	$
	m_1^*(0|a,b) = b(b+1)/((a+b)(a+b+1)),
	$
	whose supremum is approached as $a \to 0^+$, irrespective of the value of $b$. At \(x=2\),
	$
	m_1^*(2|a,b)=\{a(a+1)\}/\{(a+b)(a+b+1)\},
	$
	whose supremum is approached as \(b\to0^+\), irrespective of the value of \(a\). For $x = 1$,
	$
	m_1^*(1|a,b) = (ab)/\{(a+b)(a+b+1)\}.
	$
	The supremum of $m_1^*(1|a,b)$ is $1/4$, and it is approached precisely along sequences satisfying $
	a+b\to\infty$ and $a/(a+b)\to1/2.$
	Therefore, the limiting rule obtained by following the maximizing paths for $n=2$ is 
	\[
	\hat{\theta}_{EB}(x) =
	\begin{cases}
		0 & \text{when } x = 0 \\[4pt]
		{1}/{2} & \text{when } x = 1 \\[4pt]
		1 & \text{when } x = 2,
	\end{cases}
	\]
	\noindent which coincides with the MLE.
	
	\subsubsection{The general case of $n > 2$}
	
	We will show that $m_1^*(x\mid a,b)$ has no finite maximizer over
	$
	\mathcal H=\{(a,b):0<a<\infty,\;0<b<\infty\}.
	$
	Consider the following three cases.\\
	
	\noindent \textbf{Case 1}: $x=0$. For $x=0$, \eqref{eq:reduced_marginal_likelihood} becomes 
	\[
	m_1^*(0 \mid a,b)
	=
	\frac{\Gamma(n+b)\Gamma(a+b)}
	{\Gamma(n+a+b)\Gamma(b)}
	=
	\frac{\Gamma(n+b)}{\Gamma(b)}
	\cdot
	\frac{\Gamma(a+b)}{\Gamma(n+a+b)}
	=
	{\textstyle\prod}_{k=0}^{n-1}
	\frac{k+b}{k+a+b}
	\]
	Note that for every $k \in \{0,1,\ldots,n-1\}$, $(k+b)/(k+a+b)<1$ for all $a,b>0$. Moreover, this ratio approaches $1$ as $a\to0^+$ for fixed $b>0$, or as $b\to\infty$ for fixed $a>0$. Hence,
	$
	\sup_{a,b>0}m_1^*(0\mid a,b)=1,
	$
	and the supremum is approached as $a\to0^+$ and/or $b\to\infty$.
	However, $a=0$ makes the Beta conjugate prior improper, and hence no maximizer exists in the interior of the hyperparameter space $\mathcal{H}$.\\
	
	\noindent \textbf{Case 2}: $x=n$. Due to symmetry, a similar argument
	to that in Case 1 shows that the supremum of $m_1^*(n\mid a,b)$ is
	approached on the boundary of the hyperparameter space. Hence, no
	maximizer exists over $\mathcal{H}$.\\

	\noindent \textbf{Case 3: $1 \le x \le n-1$}. 
	Using the Beta function representation, $B(a,b)=\int_0^1 p^{a-1}(1-p)^{b-1}dp$, we rewrite \eqref{eq:reduced_marginal_likelihood} as 
	$
	m_1^*(x \mid a,b)
	= 
	\{ \int_0^1 p^{x+a-1}(1-p)^{n-x+b-1}\,dp\}
	/\{\int_0^1 p^{a-1}(1-p)^{b-1}\,dp\}.
	$
	Define 
	\begin{equation}	
		f(p)=p^x(1-p)^{\,n-x},
		\qquad 
		w_{a,b}(p)=p^{a-1}(1-p)^{b-1}.
		\label{eq:test_function_and_weight}
	\end{equation}
	Then 
	\begin{equation}
		m_1^*(x \mid a,b)
		=
		\{{\textstyle\int}_0^1 f(p)\,w_{a,b}(p)\,dp\}/
		\{{\textstyle\int}_0^1 w_{a,b}(p)\,dp\},
		\label{eq:marginal_likelihood_weighted_average}
	\end{equation}

	\noindent Note that $f$ attains a unique maximum at $p_0=x/n$, and
	$	f(p_0)=
	(x/n)^{x}
	((n-x)/n)^{(n-x)}.
	$
	
	Let $M=f(p_0)$. Then $f(p)\le M$ for all $p\in(0,1)$, which implies 
	\begin{equation}
		f(p)\,w_{a,b}(p) \le M\,w_{a,b}(p),\quad \text{since } w_{a,b}(p)>0.
		\label{eq:weighted_bound}
	\end{equation}
	Integrating both sides of the last inequality over $(0,1)$ yields
	\[
	{\textstyle\int}_0^1 f(p)\,w_{a,b}(p)\,dp
	\;\le\;
	{\textstyle\int}_0^1 M\,w_{a,b}(p)\,dp
	= M {\textstyle\int}_0^1 w_{a,b}(p)\,dp.
	\]
	Hence, from \eqref{eq:marginal_likelihood_weighted_average}:  
	$
	m_1^*(x\mid a,b)=
	(\int_0^1 f(p)\,w_{a,b}(p)\,dp)/
	(\int_0^1 w_{a,b}(p)\,dp)
	\le M.
	$	
	Thus, \(M\) is an upper bound for \(m_1^*(x\mid a,b)\).
	
	\paragraph{Result 2.1}
	\vspace{-2mm}
	For every finite $(a,b)\in\mathcal H$,
	$
	m_1^*(x\mid a,b)<M.
	$
	\paragraph{Proof of Result 2.1}
	\vspace{-2mm}
	Since $f$ is continuous (as defined at \eqref{eq:test_function_and_weight}) and has a unique maximum $M$ at $p = p_0=(x/n), p_0 \in (0,1)$,  
	there exist $\varepsilon>0$ and $\delta>0$ such that
	whenever $|p-p_0|\ge \varepsilon$, we have $ M-f(p) \ge\delta $. Therefore, $
	|p-p_0|\ge \varepsilon \;\Longrightarrow\; f(p)\le M-\delta.
	$
	
	\noindent At points close to $p_0$, from \eqref{eq:weighted_bound}: 
	\begin{equation}
		\int_{|p-p_0|<\varepsilon} f(p)\,w_{a,b}(p)\,dp
		\;\le\;
		M \int_{|p-p_0|<\varepsilon} w_{a,b}(p)\,dp.
		\label{eq:local_upper_bound}
	\end{equation}
	At points away from $p_0$, from \eqref{eq:weighted_bound}: 
	\begin{equation}
		\int_{|p-p_0|\ge\varepsilon} f(p)\,w_{a,b}(p)\,dp
		\;\le\;
		(M-\delta)\int_{|p-p_0|\ge\varepsilon} w_{a,b}(p)\,dp.
		\label{eq:outside_neighborhood_bound}
	\end{equation}
	Hence, combining \eqref{eq:local_upper_bound} and \eqref{eq:outside_neighborhood_bound}, we get 
	\[
	\begin{alignedat}{2}
		\int_0^1 f(p)\,w_{a,b}(p)\,dp
		&= 
		\int_{|p-p_0|<\varepsilon} f(p)\,w_{a,b} (p)\,dp
		&\quad +\;&
		\int_{|p-p_0|\ge\varepsilon} f(p)\,w_{a,b}(p)\,dp
		\\[6pt]
		&\le
		M\!\int_{|p-p_0|<\varepsilon} w_{a,b}(p)\,dp
		&\quad+\;&
		(M-\delta)\!\int_{|p-p_0|\ge\varepsilon} w_{a,b}(p)\,dp
	\end{alignedat}
	\]
	
	\begin{equation}
		\text{i.e.,}\quad
		\int_0^1 f(p)\,w_{a,b}(p)\,dp
		\le
		M\!\int_0^1 w_{a,b}(p)\,dp
		-
		\delta\!\int_{|p-p_0|\ge\varepsilon} w_{a,b}(p)\,dp.
		\label{eq:combined_upper_bound}
	\end{equation}
	
	\noindent Since $w_{a,b}(p)>0$ for all $p\in(0,1)$ with a finite $a$ and $b$, the second term on the right-hand side (RHS) of \eqref{eq:combined_upper_bound} is strictly positive. Therefore, 
	$
	{\textstyle\int}_0^1 f(p)\,w_{a,b}(p)\,dp
	<
	M{\textstyle\int}_0^1 w_{a,b}(p)\,dp,
	$
	hence $m_1^*(x\mid a,b)
	=
	\{{\textstyle\int}_0^1 f(p)\,w_{a,b}(p)\,dp\}/
	\{{\textstyle\int}_0^1 w_{a,b}(p)\,dp\}
	<
	M.
	$
	This proves Result 2.1.
	
	\paragraph{Result 2.2} For every $1\le x\le n-1$, let
	$
	M=(x/n)^x
	\{(n-x)/n\}^{n-x}.
	$
	A sequence $(a_s,b_s)\subset\mathcal H$ satisfies
	$
	m_1^*(x\mid a_s,b_s)\to M
	$
	iff
	$
	a_s+b_s\to\infty
	\text{ and }a_s/(a_s+b_s)\to x/n.
	$
	\paragraph{Proof of Result 2.2} See Appendix~\ref{app:proof_result32}
	\newline
	
	Combining the Results 2.1 and 2.2, we conclude that for every
	$1\le x\le (n-1)$, the function $m_1^*(x\mid a,b)$ is bounded above by
	$M=(x/n)^x\{(n-x)/n\}^{(n-x)}$, and that this bound is not attained at any finite pair $(a,b)\in\mathcal H$.
	
	\subsection{Does the MM-II Method Work for the Hyperparameters?}
	
	Note that the marginal distribution in \eqref{eq:beta_binomial_marginal} can be used to obtain
	the marginal moments of \(X\) as follows details of which are trivial
	and hence omitted for brevity:
	\begin{equation}
		E\bigl(X \mid X\sim m_1(x\mid a,b)\bigr)
		=
		n\{a/(a+b)\}.
		\label{eq:marginal_first_moment}
	\end{equation}
	Moreover,
	\begin{equation}
		E\bigl(X(X-1)\mid X\sim m_1(x\mid a,b)\bigr)
		=
		n(n-1)
		\{a/(a+b)\}
		\{(a+1)/(a+b+1)\}.
		\label{eq:marginal_second_factorial_moment}
	\end{equation}
	By equating \(X\) and \(X(X-1)\) with the right-hand sides of
	\eqref{eq:marginal_first_moment} and \eqref{eq:marginal_second_factorial_moment}, respectively, one does not get any useful
	solution (since one ends up with an inconsistent equation of "$a=-x$").
	This may not be counterintuitive since a single observation \(X\)
	cannot be used to estimate the two parameters \(a\) and \(b\).
	
	\section{Does the EBE Work with a Symmetric Beta Prior?}
	
	Section 2 showed that the ML-II method does not provide finite estimates of the two Beta hyperparameters by maximizing the marginal distribution 
	$m_1(x\mid a,b)$ (in \eqref{eq:beta_binomial_marginal}) or equivalently the proportionate term 
	$m_1^*(x\mid a,b)$ (in \eqref{eq:reduced_marginal_likelihood}). This raises the question of whether the difficulty is caused by having two unknown hyperparameters whereas the dimension of the data (or, the minimal sufficient statistic) remains 1. This motivates the study of a one-hyperparameter symmetric Beta family.
	
	Consider the symmetric Beta family, i.e., assume $a=b=c$,
	so that
	$
	\pi(\theta\mid c)
	=
	\{\theta^{c-1}(1-\theta)^{c-1}\}/B(c,c), c>0.
	$
	Substituting $a=b=c$ into
	\eqref{eq:beta_binomial_bayes_estimator} gives
	\begin{equation}
		\hat{\theta}_B
		=
		(x+c)/(n+2c).
		\label{eq:bayes_est_c}
	\end{equation}
	
	The ML-II estimate of $c$ is obtained by maximizing
	\begin{equation}
		m_1^*(x\mid c)
		=
		\{\Gamma(x+c)\Gamma(n-x+c)\Gamma(2c)\}/
		\{\Gamma(n+2c)\{\Gamma(c)\}^2\}.
		\label{eq:marginal_likelihood_symmetric}
	\end{equation}
	
	\noindent The answer depends on \(n\) and the observed value \(x\), as shown below.
	
	\subsection{The Special Case of $n=1$}
	From \eqref{eq:marginal_likelihood_symmetric},
	$
	m_1^*(x|c)
	=
	\{\Gamma(x+c)\Gamma(1-x+c)\Gamma(2c)\}/
	\{\Gamma(1+2c)\{\Gamma(c)\}^2\},
	x\in\{0,1\}.
	$
	It is easy to see that for each $x\in\{0,1\}$, the above $m_1^*(x|c)$
	is equal to \(1/2\), irrespective of the value of \(c>0\).
	Thus every \(c>0\) maximizes the marginal likelihood, and ML-II does not identify a unique value of \(c\). The corresponding posterior mean still depends on the chosen value of \(c\); it approaches the MLE only along the additional limit \(c\to0^+\).
	
	Since ML-II does not select a unique \(c\), we separately examine which fixed value \(c>0\) minimizes the maximum risk of the posterior-mean estimator in \eqref{eq:bayes_est_c}.
	It is easy to see that for any $n\ge 1$, $
	R(\hat{\theta}_B,\theta)
	=
	g(\theta|c)/(n+2c)^2,
	$
	where
	$
	g(\theta|c)
	=
	(4c^2-n)\theta^2 + (n-4c^2)\theta + c^2.
	$
	
	\noindent So, for the special case of $n=1$,
	\[
	\sup_{\theta} R_1(\hat{\theta}_B,\theta)
	=
	\begin{cases}
		1/\{4(1+2c)^2\}, & \text{if } c\le 0.5,\\[6pt]
		c^2/(1+2c)^2, & \text{if } c>0.5.
	\end{cases}
	\]
	
	The supremum risk is minimized at $c=0.5$.
	Hence
	$
	\hat{\theta}_B
	=
	(x+1/2)/2,
	$
	which coincides with the minimax estimator for the Bernoulli case $(n=1)$ with constant risk $1/16$.
	
	\subsection{The Special Case of $n=2$}
	
	For $n=2$,
	$
	m_1^*(x\mid c)
	=
	\{\Gamma(x+c)\Gamma(2-x+c)\}/
	\{2c(1+2c)\{\Gamma(c)\}^2\},
	x\in\{0,1,2\}.
	$ 
	For $x=0$,
	$
	m_1^*(0\mid c)
	=
	(1+c)/\{2(1+2c)\},
	$
	which is decreasing in $c$. Therefore
	$
	\sup_{c>0} m_1^*(0\mid c)
	=
	\lim_{c\to0^+}m_1^*(0\mid c),
	$
	and no meaningful maximizer exists.
	For $x=1$,
	$
	m_1^*(1\mid c)
	=
	(c)/\{2(1+2c)\},
	$
	which is increasing in $c$. Therefore,
	$
	\sup_{c>0} m_1^*(1\mid c)
	=
	\lim_{c\to\infty}m_1^*(1\mid c),
	$
	and again no meaningful maximizer exists.
	The case $x=2$ is similar to $x=0$.
	Hence, the posterior mean
	$
	\hat{\theta}_B
	=
	(x+c)/(2+2c)
	$
	converges to the MLE $x/2$; in other words, the ML-II does not provide a meaningful estimate of $c$.
	
	\subsection{The general case of $n>2$}
	
	Recall that
	$
	m_1^*(x|a,b)
	=
	\{\Gamma(x+a)\Gamma(n-x+b)\Gamma(a+b)\}/
	\{\Gamma(n+a+b)\Gamma(a)\Gamma(b)\}.
	$
	In the symmetric case $a=b=c$, this expression becomes
	$
	m_1^*(x|c,c)
	=
	\{\Gamma(x+c)\Gamma(n-x+c)\Gamma(2c)\}/
	\{\Gamma(n+2c)\{\Gamma(c)\}^2\}.
	$
	
	Let $\ell(c)=\ln m_1^*(x|c,c)$. If $x=0$, then
	\[
	\ell'(c)
	=
	\sum_{j=0}^{n-1}\frac{1}{c+j}
	-
	2\sum_{k=0}^{n-1}\frac{1}{2c+k}
	=
	\sum_{j=1}^{n-1}\frac{1}{c+j}
	-
	2\sum_{k=1}^{n-1}\frac{1}{2c+k}.
	\]
	For each $k=1,\ldots,n-1$,
	$
	1/(c+k)<2/(2c+k),
	$
	and therefore $\ell'(c)<0$ for all $c>0$. Hence $m_1^*(0|c,c)$ is strictly decreasing and is maximized when $c \to 0^+$, which is outside the parameter space. The case $x=n$ is identical by symmetry.
	
	It remains to consider the interior case $1\le x\le n-1$. Since $[\ln \Gamma(x)]'=\psi(x)$, where $\psi(\cdot)$ denotes the digamma function, we obtain
	$
	\ell(c)
	=
	\ln\Gamma(x+c)
	+
	\ln\Gamma(n-x+c)
	+
	\ln\Gamma(2c)
	-
	\ln\Gamma(n+2c)
	-
	2\ln\Gamma(c).
	$ Differentiating with respect to $c$ gives
	\begin{equation}\label{eq:loglik_derivative_c}
		\ell'(c)
		=
		\psi(x+c)
		+
		\psi(n-x+c)
		+
		2\psi(2c)
		-
		2\psi(n+2c)
		-
		2\psi(c).
	\end{equation}
	
	\noindent Applying the identity 
	$
	\psi(t+m)
	=
	\{\psi(t)
	+
	\sum_{i=0}^{m-1}\{1/(t+i)\}\}
	$ (Abramowitz and Stegun, 1964) to \eqref{eq:loglik_derivative_c}, we get $
	\ell'(c)
	=
	\{\psi(c+x)-\psi(c)\big\}
	+
	\{\psi(n-x+c)-\psi(c)\}
	-
	2\{\psi(n+2c)-\psi(2c)\}.
	$ Hence,
	\begin{equation}\label{eq:lprime_c_harmonic}
		\ell'(c)
		=
		\sum_{i=0}^{x-1}1/(c+i)
		+
		\sum_{j=0}^{n-x-1}1/(c+j)
		-
		2\sum_{k=0}^{n-1}1/(2c+k).
	\end{equation}
	
	\noindent Since $m_1^*(x|c,c)>0$ for all $c>0$, maximizing $m_1^*(x|c,c)$ is equivalent to maximizing $\ell(c)$. Therefore, it suffices to study the sign of $\ell'(c)$.
	From \eqref{eq:lprime_c_harmonic}, we may write
	\[
	\ell'(c)
	=
	1/c
	+
	\sum_{i=1}^{x-1}1/(c+i)
	+
	\sum_{j=1}^{n-x-1}1/(c+j)
	-
	2\sum_{k=1}^{n-1}1/(2c+k).
	\]
	
	Denote
	$
	R(c)
	=
	\sum_{i=1}^{x-1}1/(c+i)
	+
	\sum_{j=1}^{n-x-1}1/(c+j)
	-
	2\sum_{k=1}^{n-1}1/(2c+k).
	$ Thus,
	\begin{equation}\label{eq:lprime_c_decomposition}
		\ell'(c)=1/c+R(c).
	\end{equation}

	\paragraph{As $c\to0^+$.}
	The function $R(c)$ in \eqref{eq:lprime_c_decomposition}
	converges to a finite value, whereas
	$1/c\to\infty$.
	Hence,
	$\ell'(c)\to\infty$
	as $c\to0^+$.
	Therefore,
	\begin{equation}\label{eq:lprime_positive_small_c}
		\ell'(c)>0
		\qquad
		\text{for sufficiently small }c>0.
	\end{equation}
	
	\paragraph{As $c\to\infty$:}we use the Maclaurin expansion
	$
	1/(1+x)=(1-x+x^2-x^3+\cdots).
	$
	Hence
	\[
	\frac{1}{c+j}
	=
	\frac{1}{c}\, \frac{1}{1+j/c}
	=
	\frac{1}{c}\{1-\frac{j}{c}+\frac{j^2}{c^2}+\cdots\}
	=
	\frac{1}{c}-\frac{j}{c^2}+O(c^{-3}).
	\]
	
	\noindent Similarly,
	$
	1/(c+i)
	=
	1/c-i/c^2+O(c^{-3})
	$ and $
	1/(2c+k)
	=
	1/(2c)-k/(4c^2)+O(c^{-3}).
	$
	
	\noindent For the first sum:
	$
	\sum_{i=0}^{x-1}1/(c+i)
	=
	\sum_{i=0}^{x-1}\{1/c-i/c^2+O(c^{-3})\}.
	$ Therefore
	\begin{equation}\label{eq:harmonic_expansion_x}
		\sum_{i=0}^{x-1}\frac{1}{c+i}
		=
		\frac{x}{c}
		-\frac{1}{c^2}\sum_{i=0}^{x-1}i
		+O(c^{-3})
		=
		\frac{x}{c}
		-\frac{x(x-1)}{2c^2}
		+O(c^{-3}).
	\end{equation}
	
	\noindent Similarly, $
	\sum_{j=0}^{n-x-1}1/(c+j)
	=
	(n-x)/c-\{(n-x)(n-x-1)\}/(2c^2)+O(c^{-3}).
	$ For the third sum,
	$
	2\sum_{k=0}^{n-1}1/(2c+k)
	=
	2\sum_{k=0}^{n-1}
	\{
	1/(2c)-k/(4c^2)+O(c^{-3})
	\}.
	$ Hence
	\begin{equation}\label{eq:harmonic_expansion_n}
		2\sum_{k=0}^{n-1}\frac{1}{2c+k}
		=
		\frac{n}{c}
		-\frac{1}{2c^2}\sum_{k=0}^{n-1}k
		+O(c^{-3})
		=
		\frac{n}{c}
		-\frac{n(n-1)}{4c^2}
		+O(c^{-3}).
	\end{equation}
	
	\noindent Substituting \eqref{eq:harmonic_expansion_x} and \eqref{eq:harmonic_expansion_n} into \eqref{eq:lprime_c_harmonic} yields
	
	\[
	\ell'(c)
	=
	\frac{1}{c^2}
	\{
	-\frac{x(x-1)}{2}
	-\frac{(n-x)(n-x-1)}{2}
	+\frac{n(n-1)}{4}
	\}
	+
	O(c^{-3}).
	\]
	
	\noindent This expression can be written as
	$
	\ell'(c)
	=
	(1/c^2)
	\{
	n/4-(x-n/2)^2
	\}
	+
	O(c^{-3}).
	$
	
	\noindent Let
	$
	\Delta=n/4-(x-n/2)^2
	$. Then
	\begin{equation}\label{eq:lprime_asymptotic}
		\ell'(c)
		=
		\Delta/c^2
		+
		O(c^{-3}).
	\end{equation}
	
	\noindent When $\Delta\ne0$, the term $\Delta/c^2$ dominates the remainder $O(c^{-3})$, so the sign of $\ell'(c)$ in \eqref{eq:lprime_asymptotic} is determined by the sign of $\Delta$ for sufficiently large $c$.
	
	
	We are going to consider two cases as follows. Case 1: $\Delta<0$, i.e., either  $x<(n-\sqrt{n})/2$ or $x>(n+\sqrt{n})/2$; and Case 2: $\Delta \ge0$, i.e., $(n-\sqrt{n})/2 \le x \le(n+\sqrt{n})/2$. For example, if $n=16$, then Case 1 is when $x \in \{0, 1,\ldots, 5,11,12,\ldots,16\}$, Case 2 is when $x \in \{6,7,8,9,10\}$.

	\paragraph{Case 1: $\Delta<0$.}
	
	By \eqref{eq:lprime_positive_small_c},
	$\ell'(c)>0$ for sufficiently small $c>0$.
	On the other hand, by \eqref{eq:lprime_asymptotic}, the sign of
	$\ell'(c)$ is determined by the sign of $\Delta$ for sufficiently large
	$c$. Since $\Delta<0$, we have
	$\ell'(c)<0$ for sufficiently large $c$.
	Since $\ell'(c)$ is continuous, it has at least one zero in
	$(0,\infty)$. Hence, $\ell(c)$ has at least one local maximum
	at a finite value of $c$.

%

	\paragraph{Case 2: $\Delta\ge0$.}
	
	The detailed proof is given in
	Appendix~\ref{app:delta_nonnegative}.
	It follows that
	$
	\ell'(c)>0, c>0.
	$
	Hence, $\ell(c)$ is strictly increasing on $(0,\infty)$.
	Therefore, $m_1^*(x|c,c)$ has no interior maximizer, i.e., $\ell(c)$ increases monotonically  as $c \to \infty$.

	\paragraph{Numerical maximizers.}
	
	For each observation $x=0,1,\ldots,n$, let
	$\widehat c_n(x)$ denote the maximizing value of
	$m_1^*(x|c,c)$,
	where the maximum is understood as the limiting value whenever it is not attained.
	Table~\ref{tab:c_hat_theta_hat_appendix}
	in Appendix~\ref{app:numerical_maximizers}
	reports the numerically determined values of
	$\widehat c_n(x)$
	for several values of $n$ and $x$. Whenever $c \to 0^+, \hat{c}_n(x)$ is truncated at $\varepsilon = 10^{-8}$, and similarly it is taken as $10^8$ whenever $c \to \infty$ in order to maximize $m_1^*(x|c,c)$. This convention  has been used in computing $\hat{\theta}_{EB}$ and its risk $R(\hat{\theta}_{EB}, \theta)$.
	
	The following Figure~\ref{fig:logm_n10} shows the plots of $\ell(c) = \ln m_1^*(x|c)$ against $c$ with $n=10$ and for two different values of $x$ to show when $\hat{c}_n$ is finite and when it is approaching $\infty$.
	
	\begin{figure}[ht]
		\centering
		
		\begin{subfigure}[b]{0.48\textwidth}
			\centering
			\includegraphics[width=\linewidth]{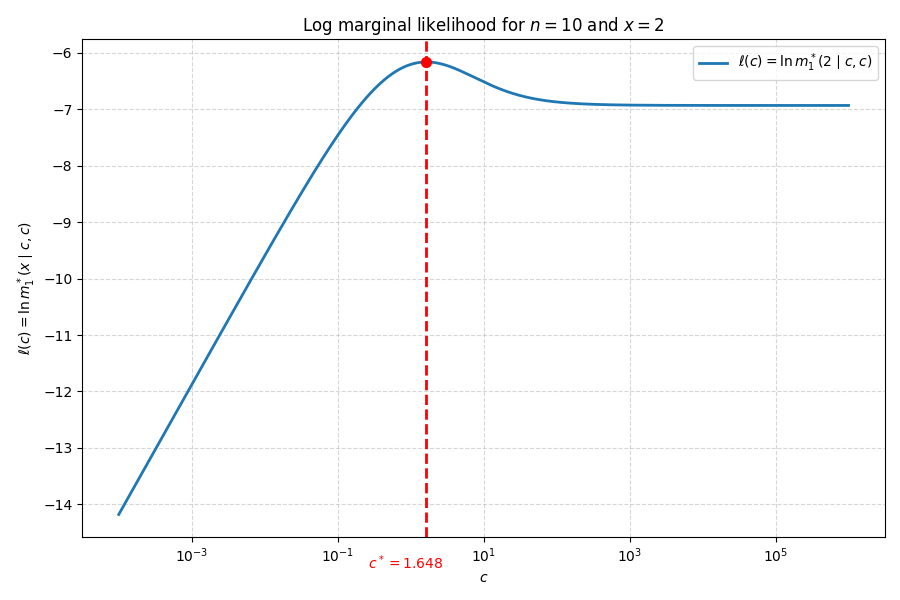}
			\caption{}
			\label{fig:logm_n10_x2}
		\end{subfigure}
		\begin{subfigure}[b]{0.48\textwidth}
			\centering
			\includegraphics[width=\linewidth]{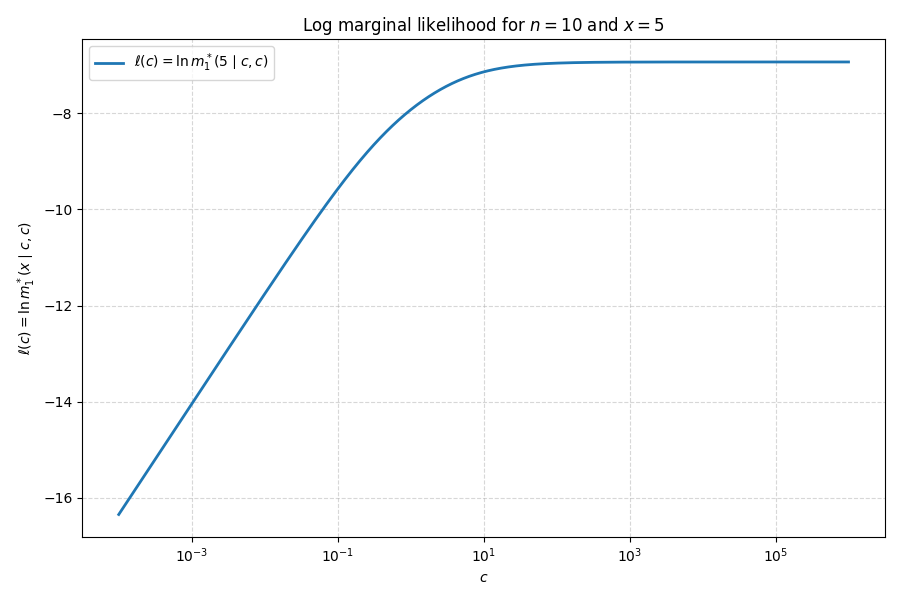}
			\caption{}
			\label{fig:logm_n10_x5}
		\end{subfigure}
		\hfill

		\caption{
			Plot of $\ell(c)=\ln m_1(x\mid c)$ w.r.t.\ $c$ when $n=10$ and
			(a) $x=2$; (b) $x=5$.
			[Note that $\widehat{c}_n=1.648$ in (a), and $=\infty$ in (b).]
		}
		\label{fig:logm_n10}
	\end{figure}

	\paragraph{Risk comparison.}
	
	We now compare the risk of the estimator obtained from the symmetric
	ML-II rule with the risk of the MLE.
	The resulting empirical Bayes estimator is
	$
	\widehat{\theta}_{EB}(x)
	=
	\{x+\widehat{c}_n(x)\}/
	\{n+2\widehat{c}_n(x)\}.
	$
	Under the squared error loss, its risk is
	\[
	R_{EB}(\theta)
	=
	R(\hat{\theta}_{EB},\theta)
	=
	\sum_{x=0}^{n}
	\left\{
	\frac{x+\widehat{c}_n(x)}
	{n+2\widehat{c}_n(x)}
	-\theta
	\right\}^{2}
	\binom{n}{x}
	\theta^x(1-\theta)^{n-x},
	\qquad 0\le\theta\le1.
	\]
	
\begin{figure}[H]
		\centering
		
		\begin{subfigure}{0.44\textwidth}
			\centering
			\includegraphics[width=\textwidth]{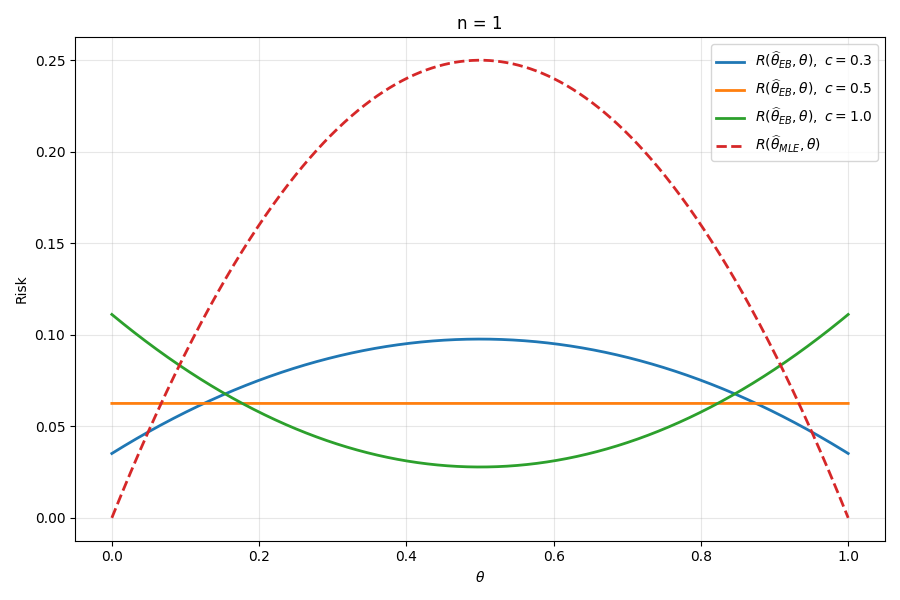}
			\caption{$n=1$}
		\end{subfigure}
		\hfill
		\begin{subfigure}{0.44\textwidth}
			\centering
			\includegraphics[width=\textwidth]{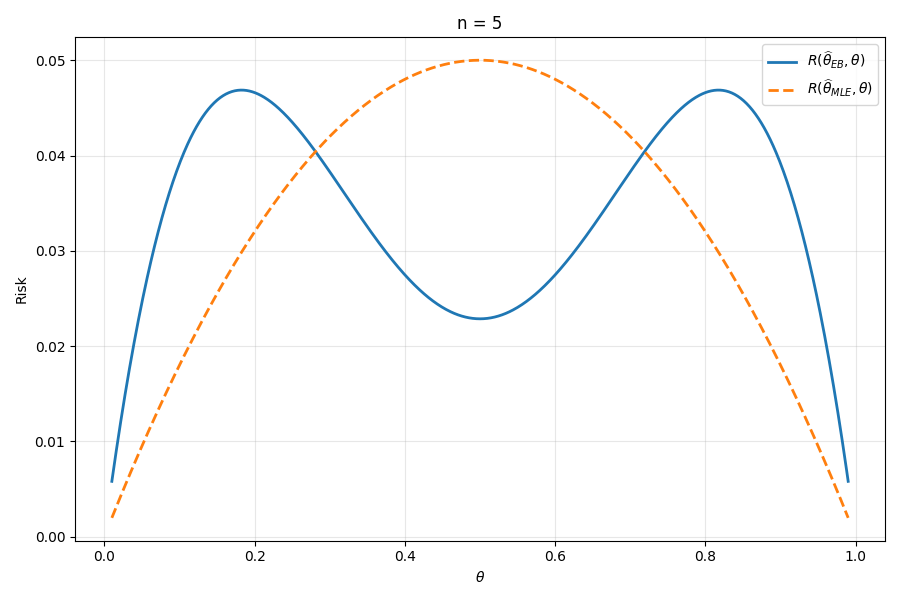}
			\caption{$n=5$}
		\end{subfigure}
		
		
		\begin{subfigure}{0.44\textwidth}
			\centering
			\includegraphics[width=\textwidth]{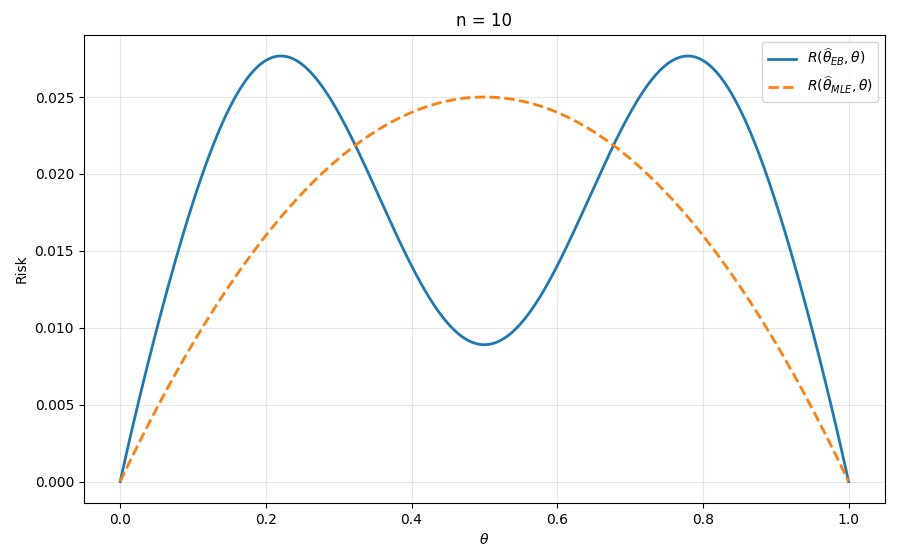}
			\caption{$n=10$}
		\end{subfigure}
		\hfill
		\begin{subfigure}{0.44\textwidth}
			\centering
			\includegraphics[width=\textwidth]{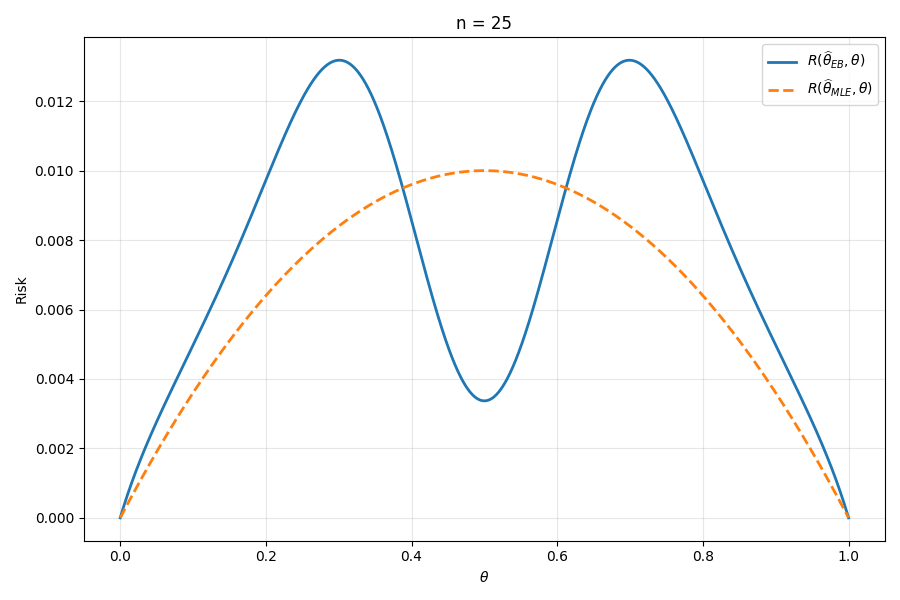}
			\caption{$n=25$}
		\end{subfigure}
		
		\caption{Risk functions of $\hat{\theta}_{EB}$ and $\hat{\theta}_{MLE}$ for $n=1,5,10$, and $25$.}
		\label{fig:risk_selected_n}
	\end{figure}

	For the MLE,
	$
	\widehat{\theta}_{MLE}(x)=x/n,
	$
	$
	R_{MLE}(\theta)
	=
	\theta(1-\theta)/n, 
	$
	and 
	$
	\sup_{\theta\in[0,1]}R_{MLE}(\theta)
	=
	1/(4n).
	$
	The two risk functions are computed numerically on
	$\theta\in[0,1]$.
	Figure~\ref{fig:risk_selected_n}
	shows the risk functions for $n=1$, $5$, $10$, and $25$.
	
	Table~\ref{tab:sup_risk_comparison} gives the approximate supremum
	risks for different values of $n$.
	
	Note that the well-known minimax estimator of $\theta$ is
	$
	\hat{\theta}_{MX}
	=
	(x+\sqrt{n}/2)/(n+\sqrt{n}),
	$
	with a constant risk of
	$
	\{4(\sqrt{n}+1)^2\}^{-1}.
	$
	So, it appears that the EBE only provides some risk improvements near
	$\theta=0.5$, and otherwise its risk is much inferior compared to that
	of the MLE.
	
	\begin{table}[H]
		\centering
		\caption{Approximate supremum risks of the empirical Bayes estimator
			and the MLE.}
		\label{tab:sup_risk_comparison}
		\begin{tabular}{ccccc}
			\hline
			$n$
			&
			$\sup R_{EB}$
			&
			$\sup R_{MLE}$
			&
			$\sup R_{EB}-\sup R_{MLE}$
			&
			Better
			\\
			\hline
			2  & 0.125000000 & 0.125000000 &  0.000000000 & Equal \\
			3  & 0.070312500 & 0.083333333 & -0.013020834 & EB \\
			4  & 0.071428569 & 0.062500000 &  0.008928569 & MLE \\
			5  & 0.046868043 & 0.050000000 & -0.003131957 & EB \\
			6  & 0.043927105 & 0.041666667 &  0.002260438 & MLE \\
			7  & 0.038909656 & 0.035714286 &  0.003195370 & MLE \\
			8  & 0.033244529 & 0.031250000 &  0.001994529 & MLE \\
			9  & 0.033443604 & 0.027777778 &  0.005665826 & MLE \\
			10 & 0.027666499 & 0.025000000 &  0.002666499 & MLE \\
			16 & 0.019680138 & 0.015625000 &  0.004055138 & MLE \\
			32 & 0.010401879 & 0.007812500 &  0.002589379 & MLE \\
			64 & 0.005508976 & 0.003906250 &  0.001602726 & MLE \\
			\hline
		\end{tabular}
	\end{table}
	
	The comparison depends on $n$.
	For $n=3$ and $n=5$, the empirical Bayes estimator has a smaller
	supremum risk than the MLE.
	For $n=2$, the two supremum risks are equal.
	For $n=4$ and for all values $n\ge6$ considered here,
	the MLE has a smaller supremum risk.
	Thus, the empirical Bayes estimator is not always better than the MLE.
	The existence of a finite local maximizer of
	$m_1^*(x\mid c,c)$ does not ensure a smaller worst-case risk.
	
	\paragraph{Remark 3.1.}
	Can we use the MM-II approach to estimate the single hyperparameter
	\(c\) in the symmetric Beta prior case?
	Note that with \(a=b=c\), \eqref{eq:marginal_first_moment} yields \(E(X)=n/2\), which is
	of no use. Also, \eqref{eq:marginal_second_factorial_moment} yields
	$
	E(X(X-1))
	=
	(n/2)(n-1)\{(c+1)/(2c+1)\}.
	$
	If the expectant \(X(X-1)\) is equated with the right-hand side, then two options
	can be followed: (i) replace \(X\) by \(n/2\), since the first moment
	expression says so; or (ii) try to solve for \(c\) using the whole
	expression
	$
	X(X-1)
	=
	(n/2)(n-1)\{(c+1)/(2c+1)\}.
	$
	The above first option yields
	$
	c=-(n-X)/(n-2X+1);
	$
	and the second option yields
	$
	c=
	\{(n/2)(n-1)-X(X-1)\}/
	\{2X(X-1)-(n/2)(n-1)\}.
	$
	Both of these solutions are infeasible as they can take negative
	values over a substantial part of the sample space.

	\medskip

	
	\section{Risk Comparison and Supremum Risk}
	
	Section~3.3 shows that the normalized marginal likelihood $m_1^*(x|a,b)$
	does not attain its supremum at any finite $(a,b)$, although the supremum can be
	approached along suitable sequences. This phenomenon does not determine the frequentist risk performance of the posterior-mean estimators with fixed hyperparameters. In particular, the absence of a marginal-likelihood maximizer does not preclude the existence of fixed $(a,b)$ for which the corresponding posterior-mean estimator has smaller supremum risk than the MLE.
	
	We characterize the set of $(a,b)$ such that
	$
	\sup_{\theta\in[0,1]} R_{a,b}(\theta)
	<
	\sup_{\theta\in[0,1]} R_{MLE}.
	$
	Recall that, under squared error loss,
	$
	R_{MLE}(\theta)=\theta(1-\theta)/n\Rightarrow
	\sup_{\theta\in[0,1]}R_{MLE}(\theta)=1/4n.
	$
	
	For the posterior-mean estimator
	$
	\hat{\theta}_{a,b}(X)=(X+a)/(n+a+b), a,b>0,
	$
	with fixed \(a,b\), the risk is
	$
	R_{a,b}(\theta)
	=
	\{n\theta(1-\theta)+\big(a(1-\theta)-b\theta\big)^2\}/(n+a+b)^2.
	$
	
	Let $
	g(\theta)=n\theta(1-\theta)+\big(a(1-\theta)-b\theta\big)^2.
	$
	Then
	$
	\sup_{\theta\in[0,1]}R_{a,b}(\theta)=\{\sup_{\theta\in[0,1]}g(\theta)\}/\{(n+a+b)^2\}.
	$
	A straightforward expansion yields
	$
	g(\theta)=A\theta^2+B\theta+C,
	$
	where
	$
	A=(a+b)^2-n, B=n-2a(a+b), C=a^2.
	$
	Hence the behavior of $\sup_{\theta\in[0,1]}g(\theta)$ is governed by the sign of $A$:	
	If $A\ge 0$, then $g(\theta)$ is convex (or linear when $A=0$), so $\sup_{\theta\in[0,1]}g(\theta)=\max\{g(0),g(1)\}=\max\{a^2,b^2\}.$ 	
	If $A<0$, then $g(\theta)$ is concave and the interior critical point
	$
	\theta^*=-B/(2A)=\{n-2a(a+b)\}/\{2\{\,n-(a+b)^2\,\}\}
	$
	is the maximizer iff $\theta^*\in[0,1]$; otherwise the maximum is attained at a boundary point. Therefore, the evaluation of $\sup_{\theta\in[0,1]} R_{a,b}(\theta)$ reduces to a case-by-case analysis according to the sign of $A$. We consider the corresponding cases below.
	
	
	\subsubsection*{Case $A\ge 0$ (i.e., $(a+b)^2\ge n$)}
	
	In this case,
	$
	\sup_{\theta\in[0,1]}R_{a,b}(\theta)=\{\max\{a^2,b^2\}\}/(n+a+b)^2.
	$
	Thus $\sup R_{a,b}<\sup R_{MLE}$ is equivalent to
	\begin{equation}
		\{\max\{a^2,b^2\}\}/(n+a+b)^2<1/(4n)
		\quad\Longleftrightarrow\quad
		2\sqrt{n}\,\max\{a,b\}<n+a+b.
		\label{eq:risk_improvement_condition}
	\end{equation}
	
	\paragraph{Subcase $b\ge a$.}
	Then $\max\{a,b\}=b$ and \eqref{eq:risk_improvement_condition} becomes
	$2\sqrt{n}\,b<(n+a+b)$, or equivalently,
	$(2\sqrt{n}-1)b<(n+a)$.
	Hence the feasible region in this subcase is
	\begin{equation}
		\left\{
		\begin{aligned}
			&(a+b)^2\ge n,\\
			&b\ge a,\\
			&(2\sqrt{n}-1)b<n+a.
		\end{aligned}
		\right.
		\label{eq:sufficient_conditions}
	\end{equation}
	
	\paragraph{Subcase $a\ge b$.}
	Similarly, $\max\{a,b\}=a$ and \eqref{eq:risk_improvement_condition} yields
	$
	2\sqrt{n}\,a<(n+a+b)$, or equivalently, 
	$
	(2\sqrt{n}-1)a<(n+b),
	$
	so the feasible region is described by
	\begin{equation}
		\left\{
		\begin{aligned}
			&(a+b)^2\ge n,\\
			&a\ge b,\\
			&(2\sqrt{n}-1)a<n+b.
		\end{aligned}
		\right.
		\label{eq:risk_improvement_sufficient_conditions}
	\end{equation}
	
	To show that the feasible region described by \eqref{eq:sufficient_conditions}–\eqref{eq:risk_improvement_sufficient_conditions}
	is non-empty, it suffices to exhibit at least one pair $(a,b)$ satisfying
	these inequalities.
	
	Observe that the symmetric choice
	$a=b=\sqrt{n}/2$ 
	satisfies $(a+b)^2 = n$ and
	$
	(2\sqrt{n}-1)a
	=
	(2\sqrt{n}-1)b
	<
	(n+a),
	$
	so that both \eqref{eq:sufficient_conditions} and \eqref{eq:risk_improvement_sufficient_conditions} hold.
	Hence the feasible region in the case $A\ge 0$ is non-empty.
	
\begin{figure}[H]
		\centering
		
		\begin{subfigure}{0.48\textwidth}
			\centering
			\includegraphics[width=\linewidth]{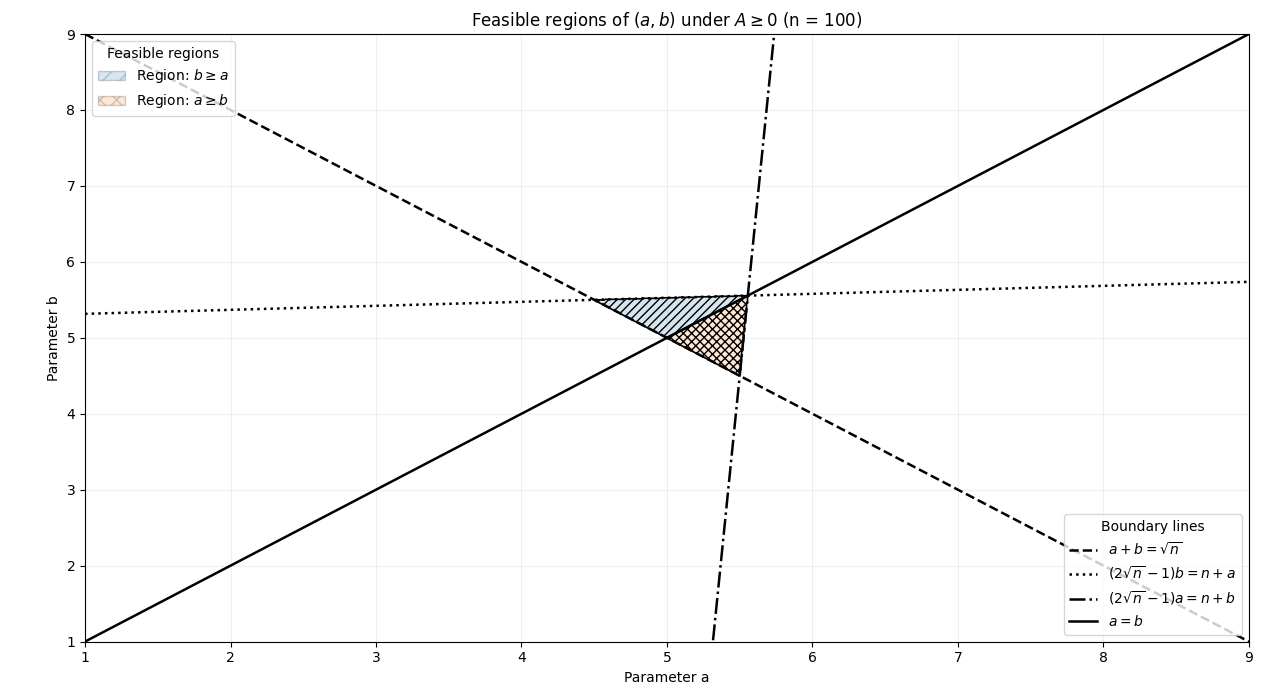}
			\caption{$n=100$.}
			\label{fig:fig1a_A_ge_0_n100_binomial}
		\end{subfigure}
		\hfill
		\begin{subfigure}{0.48\textwidth}
			\centering
			\includegraphics[width=\linewidth]{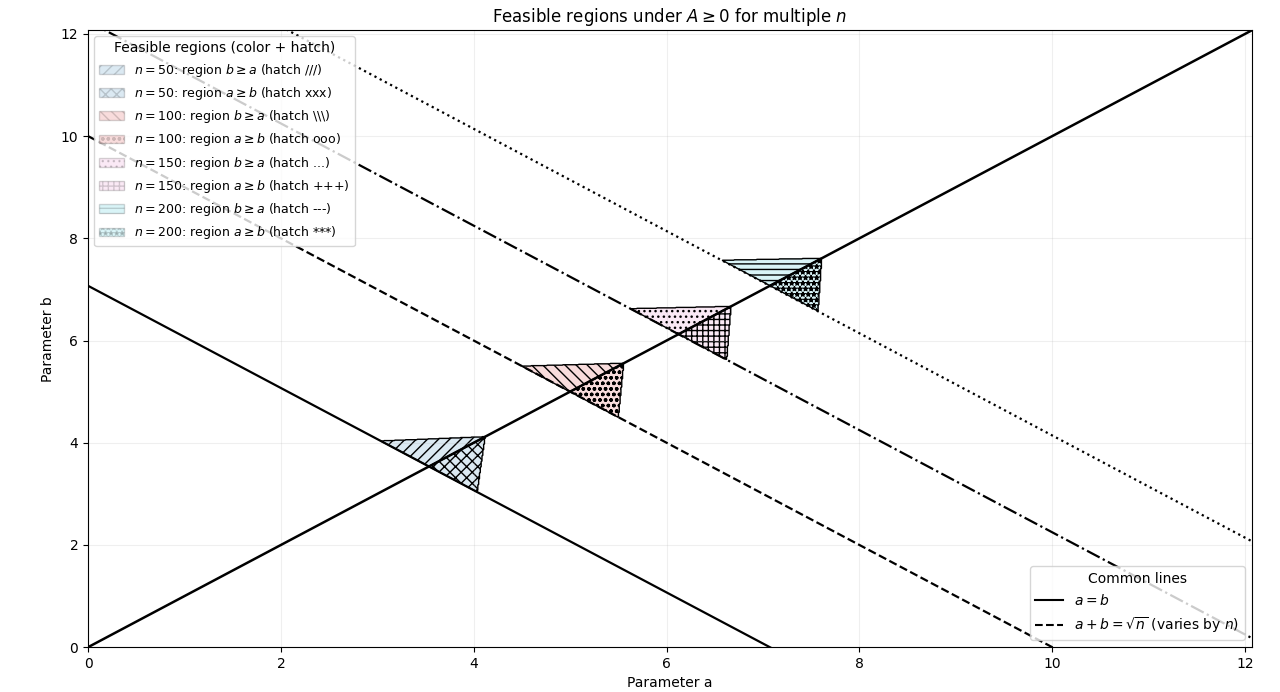}
			\caption{Multiple values of $n$.}
			\label{fig:fig1b_A_ge_0_multi_values_of_n_binomial}
		\end{subfigure}
		
		\caption{Feasible regions in the $(a,b)$-plane where 
			$\sup_{\theta\in[0,1]} R_{a,b}(\theta)
			< \sup_{\theta\in[0,1]} R_{MLE}(\theta)$ under $A \ge 0$.}
		\label{fig:A_ge_0_combined}
	\end{figure}

	
	Figure~\ref{fig:A_ge_0_combined} illustrates the feasible regions
	in the $(a,b)$-plane where
	$
	\sup_{\theta\in[0,1]} R_{a,b}(\theta)
	<
	\sup_{\theta\in[0,1]} R_{MLE}
	$
	under the case $A \ge 0$.
	Panel~(a) corresponds to $n=100$, while panel~(b) displays the regions
	for multiple values of $n$.
	
	For each fixed $n$, the admissible set consists of two symmetric triangular
	domains separated by the line $a=b$. As shown in panel~(b), the same geometric
	structure is preserved when $n$ varies, while the vertices scale proportionally
	with $\sqrt{n}$. This suggests that the region in which the posterior-mean estimator improves upon the MLE in sup-risk is concentrated around the minimax configuration
	$a=b=\sqrt{n}/2$.
	
	\subsubsection*{Case $A<0$ (i.e., $(a+b)^2<n$)}
	
	When $A<0$, the maximizer depends on whether $\theta^*\in[0,1]$.
	
	\paragraph{Subcase 1: $\theta^*\in(0,1)$.}
	From the previous derivation,
	$
	\theta^*\in(0,1) \Longleftrightarrow 
	a(a+b) < n/2\ \ \text{and}\ \ b(a+b) <n/2,
	$
	together with $(a+b)^2<n$.
	In this subcase,
	\[
	\sup_{\theta\in[0,1]}R_{a,b}(\theta)={g(\theta^*)}/{(n+a+b)^2},
	\qquad
	g(\theta^*)=a^2+{\big(n-2a(a+b)\big)^2}/{4\{\,n-(a+b)^2\,\}}.
	\]
	Therefore, $\sup R_{a,b}<1/(4n)$ is equivalent to
	\[
	\left\{
	\begin{aligned}
		&(a+b)^2<n,\\
		&a(a+b) < n/2,\\
		&b(a+b) < n/2,\\
		&4a^2n+{n\{n-2a(a+b)\}^2}/\{\,n-(a+b)^2\,\}<(n+a+b)^2.
	\end{aligned}
	\right.
	\]
	
	\paragraph{Subcase 2: maximizer at $\theta=0$.}
	If $a(a+b) \ge n/2$ (so $\theta^*\le 0$), then
	$
	\sup_{\theta\in[0,1]}R_{a,b}(\theta)=a^2/(n+a+b)^2.
	$
	The condition $\sup R_{a,b}<1/(4n)$ becomes
	$
	a^2/(n+a+b)^2<1/(4n)$
	or equivalently, 
	$
	b>2a\sqrt{n}-n-a.
	$
	Thus the feasible region is described by
	\[
	\left\{
	\begin{aligned}
		&(a+b)^2<n,\\
		&a(a+b)\ge n/2,\\
		&b>2a\sqrt{n}-n-a.
	\end{aligned}
	\right.
	\]
	
	\paragraph{Subcase 3: maximizer at $\theta=1$.}
	If $b(a+b) \ge n/2$ (so $\theta^*\ge 1$), then
	$
	\sup_{\theta\in[0,1]}R_{a,b}(\theta)=b^2/(n+a+b)^2,
	$
	and $\sup R_{a,b}<1/(4n)$ is equivalent to
	$
	n+a+b>2\sqrt{n}\,b.
	$
	Hence the feasible region is
	\[
	\left\{
	\begin{aligned}
		&(a+b)^2<n,\\
		&b(a+b)\ge n/2,\\
		&n+a+b>2\sqrt{n}\,b.
	\end{aligned}
	\right.
	\]
	
	Figure~\ref{fig:A_less_than_0_combined} below illustrates the feasible region under $A<0$, i.e., $(a+b)^2<n$, where the condition $\sup R_{a,b} < \sup R_{MLE}$ holds.
	\begin{figure}[H]
		\centering
		
		\begin{subfigure}{0.48\textwidth}
			\centering
			\includegraphics[width=\linewidth]{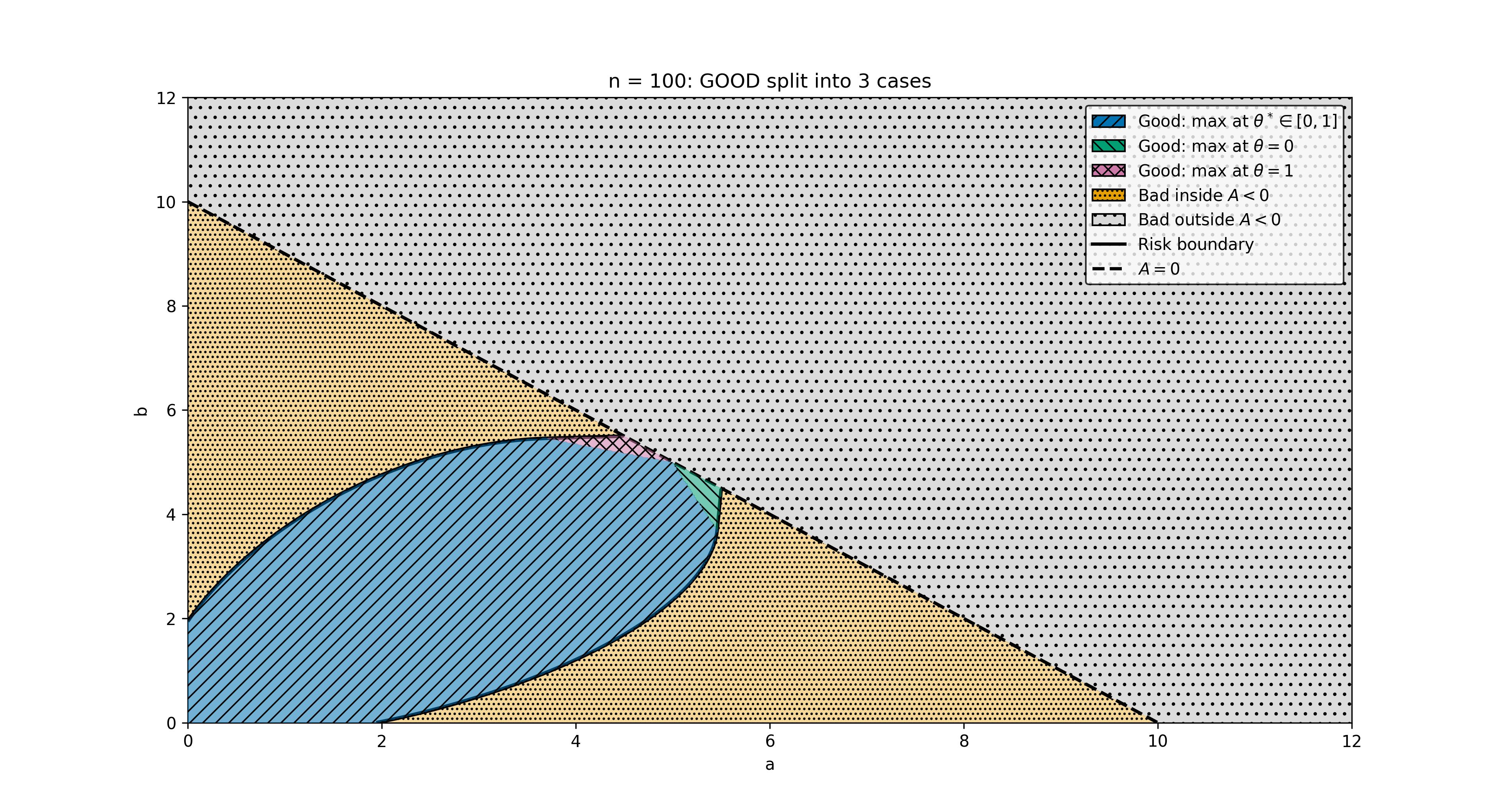}
			\caption{$n=100$.}
			\label{fig:fig2a_A_less_than_0_n100_binomial}
		\end{subfigure}
		\hfill
		\begin{subfigure}{0.48\textwidth}
			\centering
			\includegraphics[width=\linewidth]{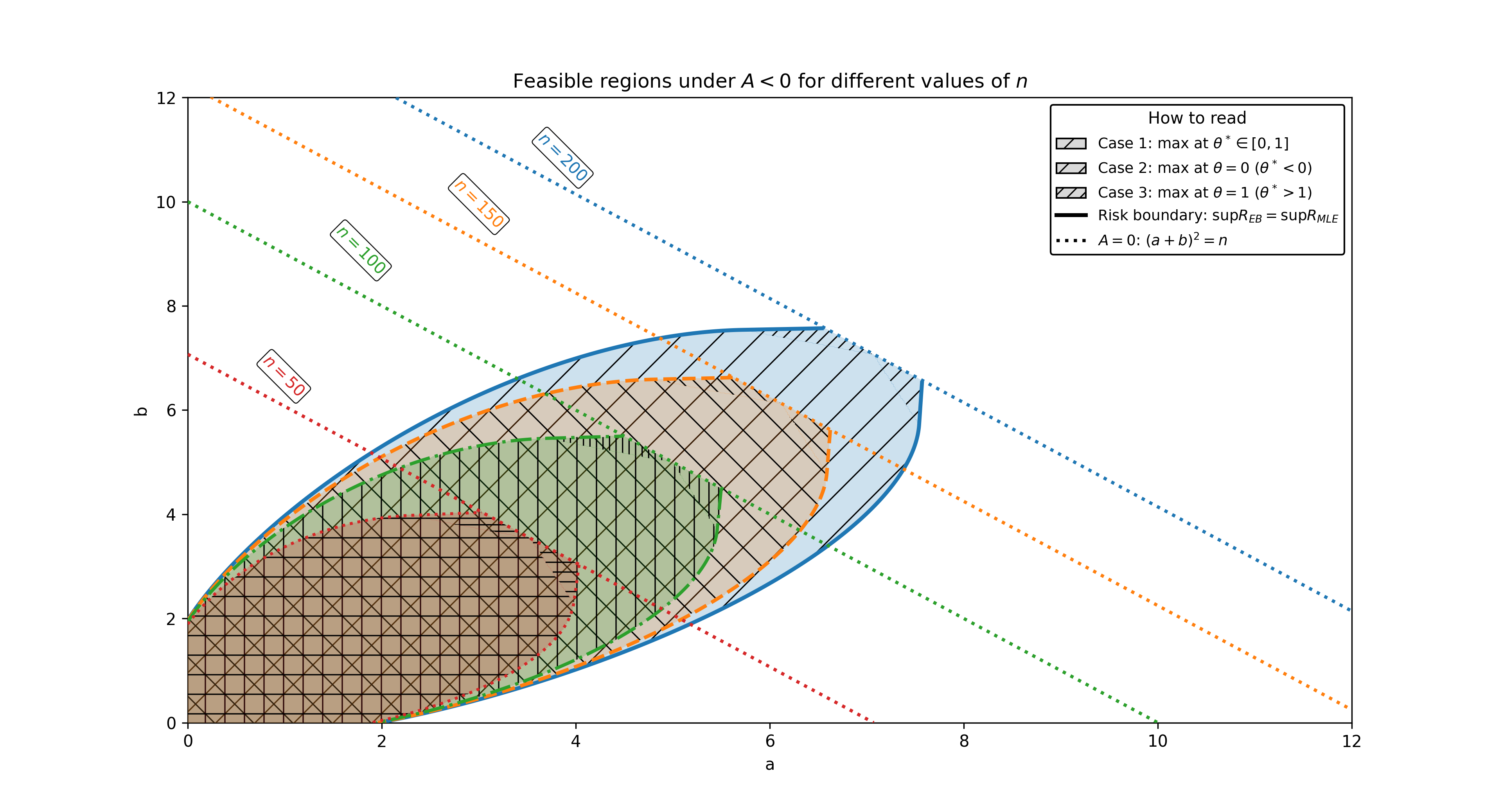}
			\caption{Multiple values of $n$.}
			\label{fig:fig2b_A_less_than_0_multi_n_binomial}
		\end{subfigure}
		
		\caption{Feasible regions under $A<0$, i.e., $(a+b)^2<n$.}
		\label{fig:A_less_than_0_combined}
	\end{figure}
	
	Under $A<0$, the maximizer of $R_{a,b}$ may occur either at an interior point 
	$\theta^*\in(0,1)$ or at the boundary $\theta=0$ or $\theta=1$. 
	The shaded region represents the set of $(a,b)$ for which 
	$\sup R_{a,b} < \sup R_{MLE}$. 
	Panel (b) shows how the feasible region changes as $n$ varies.
	
	The previous analysis shows that there exist choices of $(a,b)$ for which 
	$
	\sup_{\theta} R_{a,b}(\theta) < \sup_{\theta} R_{MLE}(\theta).
	$

	\subsubsection*{Binomial under an Alternative Weighted Loss}
	
	The previous subsection considered the usual squared error loss and compared the MLE with posterior-mean estimators having fixed hyperparameters. We consider the same question under a loss function that places heavier weight near $\theta=0$ and $\theta=1$.
	
	Let $X\sim \mathrm{Bin}(n,\theta)$ with $0<\theta<1$, and consider the weighted loss
	$
	L(\theta,\hat\theta)=(\hat\theta-\theta)^2/\{\theta(1-\theta)\}.
	$ For the MLE $\hat\theta_{MLE}(X)=X/n$, 
	$
	R(\theta,\hat\theta_{MLE})=1/n,
	$
	which is constant in $\theta$.
	Next, consider the posterior mean estimator under the beta prior $\mathrm{Beta}(a,b)$,
	$
	\hat\theta_B(X)=(X+a)/(n+a+b), a>0,\; b>0.
	$ 
	A direct calculation yields
	
	\begin{equation}\label{eq:weighted_risk}
		R(\theta,\hat\theta_B)
		=
		n/(n+a+b)^2
		+
		\{a-(a+b)\theta\}^2/
		\{(n+a+b)^2\,\theta(1-\theta)\}.
	\end{equation}
	
	As $\theta\to 0^+$, \eqref{eq:weighted_risk} yields
	$
	R(\theta,\hat\theta_B)\sim a^2/\{(n+a+b)^2\,\theta\}\to\infty,
	$
	and as $\theta\to 1^-$,
	$
	R(\theta,\hat\theta_B)\sim b^2/\{(n+a+b)^2\,(1-\theta)\}\to\infty.
	$
	Hence,
	$
	\sup_{0<\theta<1} R(\theta,\hat\theta_B)=\infty.
	$
	In contrast,
	$
	\sup_{0<\theta<1}R(\theta,\hat\theta_{MLE})
	=
	1/n
	$. Therefore, under the weighted loss
	$
	L(\theta,\hat\theta)=(\hat\theta-\theta)^2/\{\theta(1-\theta)\},
	$
	every posterior-mean estimator with fixed \(a,b>0\) has infinite worst-case risk, and no meaningful minimax comparison with the MLE is possible within this family.
	
	Under the standard squared error loss, under which both the MLE and the posterior-mean estimators with fixed hyperparameters have finite and comparable risk functions. Under the weighted loss above,
	$
	R(\theta,\hat\theta_B)\to\infty
	$
	as
	$
	\theta\to0^+
	$
	or
	$
	\theta\to1^-.
	$	
	Thus, the risk comparison depends strongly on the choice of loss function.
	
	\section{Extension to the Poisson Model}
	
	Consider a random sample $
	X_1,\ldots,X_n \stackrel{\mathrm{iid}}{\sim} \mathrm{Poisson}(\theta), \theta>0,$
	and let $T=\sum_{i=1}^n X_i$ .
	Since sums of independent Poisson random variables are again Poisson, $T\sim \mathrm{Poisson}(n\theta)$. We evaluate estimators under the weighted squared error loss
	$
	L(\theta,\hat\theta)=(\hat\theta-\theta)^2/\theta,
	\theta>0.
	$
	This loss reflects the fact that
	$
	\mathrm{Var}_\theta(X_i)=\theta.
	$
	
	\subsection{Risk under the weighted loss}
	
	The MLE is
	$\hat\theta_{MLE}(T)= T/n$. Since $E_\theta\,(T/n)=\theta$ and
	$\mathrm{Var}_\theta\,(T/n) =\theta/n$, 
	we have
	$
	R(\theta,\hat\theta_{MLE})
	=
	E_\theta\,[((T/n)-\theta)^2/\theta]
	=
	(1/\theta)\,\mathrm{Var}_\theta(T/n)
	=
	1/n.
	$
	Thus the MLE has constant risk equal to \(1/n\).
	
	Now consider a general affine estimator
	$
	\hat\theta(T)=\alpha T+\beta,
	\alpha,\beta\in\mathbb{R}.
	$

	Using \(E_\theta(T)=n\theta\) and \(\mathrm{Var}_\theta(T)=n\theta\), we get
	$
	R(\theta,\hat\theta)
	=
	E_\theta\!\left[(\alpha T+\beta-\theta)^2/\theta\right]
	=
	\alpha^2\mathrm{Var}_\theta(T)/\theta
	+
	\{\alpha E_\theta(T)+\beta-\theta\}^2/\theta.
	$ 
	Hence
	$
	R(\theta,\hat\theta)
	=
	\alpha^2 n
	+
	\{(\alpha n-1)\theta+\beta\}^2/\theta.
	$ 
	Expanding the last term gives
	$
	R(\theta,\hat\theta)
	=
	\alpha^2 n
	+
	(\alpha n-1)^2\theta
	+
	2(\alpha n-1)\beta
	+
	\beta^2/\theta.
	$
	
	Consider the limits as
	$
	\theta\to\infty
	$
	and
	$
	\theta\to0^+
	$. 
	If $
	\alpha n\neq 1$, then the term \((\alpha n-1)^2\theta\) forces $R(\theta,\hat\theta)\to\infty
	\text{ as }\theta\to\infty$. If $\beta\neq 0$, then the term \(\beta^2/\theta\) forces $R(\theta,\hat\theta)\to\infty \text{ as }\theta\to 0^+$.
	Therefore, the only affine estimator with finite worst-case risk is obtained by choosing $\alpha=(1/n), \beta=0$, which is exactly the MLE. In particular, $
	\sup_{\theta>0} R(\theta,\hat\theta) = \infty$ for every affine estimator $\hat\theta(T)\neq T/n$.
	
	\subsection{Consequences for Poisson--Gamma posterior-mean estimators}
	
	Now assume the conjugate prior $
	\theta\sim \mathrm{Gamma}(a,b), a>0,\; b>0,
	$
	with density
	$
	\pi(\theta\mid a,b)=\{b^a/\Gamma(a)\}\theta^{a-1}e^{-b\theta},
	\theta>0,
	$
	where \(b\) is the rate parameter. Then the posterior distribution is $\mathrm{Gamma}(T+a,b+n)$, so the posterior mean is 
	$
	\hat\theta_B(T)=E(\theta\mid T)=(T+a)/(b+n).
	$
	Thus,
	$
	\hat\theta_B(T)=\alpha T+\beta,
	\alpha=1/(b+n),
	\beta=a/(b+n).
	$
	Here we evaluate this posterior-mean estimator under the weighted loss,
	it is not the Bayes rule under that loss.
	Since $\hat\theta_B(T)\neq (T/n)$ for any finite \(a>0\) and \(b>0\), the previous result already implies
	$
	\sup_{\theta>0}R(\theta,\hat\theta_B)=\infty.
	$	
	Therefore, under the loss $
	L(\theta,\hat\theta)=(\hat\theta-\theta)^2/\theta$,
	the Poisson--Gamma posterior-mean estimator has infinite worst-case risk, whereas the MLE has the constant risk \(1/n\).
	
%
	\subsection{Failure of the ML-II approach in the Poisson--Gamma family}
	
	We consider the existence of ML-II estimates for the Gamma hyperparameters in the Poisson model. Since
	$
	T\mid\theta\sim\mathrm{Poisson}(n\theta),
	$
	the marginal (prior predictive) distribution of \(T\) is obtained by integrating the likelihood with respect to the Gamma prior (Gelman et al. (2013)):
	\[
	m_2(t\mid a,b)
	=
	\int_0^\infty
	\frac{e^{-n\theta}(n\theta)^t}{t!}\,
	\frac{b^a}{\Gamma(a)}
	\theta^{a-1}e^{-b\theta}\,d\theta.
	\]
	Evaluating the integral gives
	\[
	m_2(t\mid a,b)
	=
	\frac{n^t b^a}{\Gamma(a)}
	\frac{\Gamma(t+a)}{t!(b+n)^{t+a}}.
	\]
	Since \(n^t/t!\) does not depend on \(a\) or \(b\), maximizing \(m_2(t\mid a,b)\) is equivalent to maximizing
	\[
	m_2^*(t\mid a,b)
	=
	\frac{b^a\,\Gamma(t+a)}{\Gamma(a)(b+n)^{t+a}},
	\qquad a>0,\; b>0.
	\]
	
	\noindent \textbf{Case \(t=0\)}: When \(t=0\),
	$
	m_2^*(0\mid a,b)
	=
	(b/(b+n))^a.
	$
	For all \(a,b>0\),
	$
	0<m_2^*(0\mid a,b)<1.
	$
	Moreover, for fixed \(a>0\),
	$
	m_2^*(0\mid a,b)\to 1
	\text{ as }b\to\infty,
	$
	and for fixed \(b>0\),
	$
	m_2^*(0\mid a,b)\to 1
	\text{ as }a\to 0^+.
	$
	Hence
	$
	\sup_{a,b>0}m_2^*(0\mid a,b)=1,
	$
	but this value is never attained at any finite \((a,b)\). Therefore, no ML-II estimator exists when \(t=0\).
	
	\noindent \textbf{Case \(t\ge 1\)}: Now fix \(t\ge 1\) and define
	\[
	\ell(a,b)=\log m_2^*(t\mid a,b)
	=
	a\log b+\log\Gamma(t+a)-\log\Gamma(a)-(t+a)\log(b+n).
	\]
	For fixed \(a\),
	$
	\partial \ell/\partial b
	=
	a/b-(t+a)/(b+n).
	$
	Let \(b^*(a)\) denote the critical point of \(\ell(a,b)\) with respect to \(b\). Setting \(\partial\ell/\partial b=0\) gives \(b^*(a)=na/t\). Since
	$
	\partial\ell/\partial b
	=
	(an-tb)/\{b(b+n)\},
	$
	and \(b(b+n)>0\), $\partial\ell/\partial b$ is positive for \(b<na/t\) and negative for \(b>na/t\). Hence, \(b^*(a)\) is the unique global maximizer of \(\ell(a,b)\) for each fixed \(a>0\).
	Substituting \(b=b^*(a)\) into \(\ell(a,b)\) gives
	$
	g(a)=\ell(a,na/t).
	$
	Differentiating \(g(a)\), we obtain
	$
	g'(a)
	=
	\log (a/t)
	+\psi(t+a)-\psi(a)
	-\log\!\left(1+a/t\right),
	$
	where \(\psi\) is the digamma function. Since
	$
	\log (a/t)-\log\!\left(1+a/t\right)
	=
	-\log\!\left(1+t/a\right),
	$
	this becomes
	$
	g'(a)=\psi(t+a)-\psi(a)-\log\!\left(1+t/a\right).
	$
	Using the identity for the digamma function (Abramowitz and Stegun (1964)), 
	$
	\psi(t+a)-\psi(a)=\sum_{j=0}^{t-1}1/(a+j),
	$
	we get
	$
	g'(a)
	=
	\sum_{j=0}^{t-1}1/(a+j)
	-
	\log\!\left(1+t/a\right).
	$
	Now,
	$
	\sum_{j=0}^{t-1}1/(a+j)
	>
	\int_a^{a+t}(du/u)
	=
	\log\!\left(1+t/a\right),
	$
	so
	$
	g'(a)>0$ for all $a>0$.
	Therefore, the supremum is approached as
	\(a\to\infty\) with \(b=b^*(a)=na/t\)
	(and hence \(b\to\infty\)), but it is not attained at any finite
	pair \((a,b)\).
	Hence, for every \(t\ge 1\), the ML-II maximizer does not exist in the Poisson--Gamma family.
	
	\subsection{Conclusion for the Poisson case}
	
	The Poisson results lead to two conclusions. First, under the loss
	$
	L(\theta,\hat\theta)=(\hat\theta-\theta)^2/\theta.
	$
	the MLE has constant risk \(1/n\), whereas every Poisson--Gamma posterior-mean estimator with fixed finite \(a,b>0\) has infinite worst-case risk. In fact, the MLE is the only affine estimator with finite worst-case risk. Second, the ML-II procedure does not provide finite estimates of \(a,b\), because the marginal likelihood has no maximizer in the admissible hyperparameter space.
	
	The weighted-loss result is similar to the corresponding Binomial result: in both models, the MLE has constant risk, while posterior-mean estimators with fixed positive hyperparameters have infinite worst-case risk. The Poisson result extends this conclusion: every affine estimator other than the MLE has infinite worst-case risk.

	\section{Concluding Remark:}	
	
	The Binomial example shows that the ML-II approach does not provide any meaningful estimate(s) of the Beta
	hyperparameter(s) considered here. For boundary observations, the marginal likelihood of the data keeps monotonically
	increasing as the hyperparameter(s) keep(s) moving toward the boundary of the parameter space. For observations
	well inside the range, the supremum of the marginal distribution is approached along sequence of the hyperparameter values
	in such a way that the EBE coincides with the traditional MLE. A very similar result holds for the Poisson model with a 
	Gamma family of priors. Thus, the failure of ML-II method in producing a meaningful EBE (which is different from the 
	traditional MLE) is not restricted to the Binomial--Beta combination, and may extend to many other problems as well.
	However, the failure of the ML-II step does not imply that the Bayes estimator in the form of posterior mean is 
	ineffective. For the Binomial model, it is possible to choose the hyperparameter(s) so that the Bayes estimator can 
	have a smaller maximum risk than the MLE under squared error loss, and the minimax estimator also belongs to this 
	collection of possible hyperparameter(s). Thus, the difficulty comes from the ML-II approach, not from the posterior-mean 
	estimator itself.
	Future work may study whether similar non-existence of EBE could hold for other conjugate priors with more 
	hyperparameters as well as different loss functions. The work presented here actually stemmed from our pursuit to
	study the EBE of common mean of several Normal distributions with unknown and possibly unequal variances with
	suitable conjugate family of priors. So far, the preliminary numerical trends indicate that the same patterns observed
	in the case of Binomial are still at play here, but the details will be shared in a future formal report.

	\section*{Acknowledgements}
	We acknowledge Ho Chi Minh City University of Technology (HCMUT), VNU-HCM for supporting this study.

	\appendix
	\section{Appendix}
	
	\subsection{Proof of Result 2.2}
	\label{app:proof_result32}
	It is easy to see that
	$
	\Gamma(a+n)=\Gamma(a)\prod_{j=0}^{n-1}(a+j),
	$
	which, upon applying to \eqref{eq:reduced_marginal_likelihood}, yields
	\begin{align}
		m_1^*(x\mid a,b)
		&=
		\{
		{\textstyle\prod}_{j=0}^{x-1}(a+j)
		{\textstyle\prod}_{k=0}^{n-x-1}(b+k)
		\}
		/
		\{
		{\textstyle\prod}_{m=0}^{n-1}(a+b+m)\}.
		\label{eq:gamma_product_expansion}
	\end{align}
	
	Now consider the denominator of \eqref{eq:gamma_product_expansion} first, which can be written as 
	\[
	{\textstyle\prod}_{m=0}^{n-1}(a+b+m)
	=
	\{{\textstyle\prod}_{m=0}^{x-1}(a+b+m)\}
	\{{\textstyle\prod}_{\ell=x}^{n-1}(a+b+\ell)\}.
	\] 
	
	Let $\ell=(x+\ell')$, so that $\ell'$ runs from $0$ to $(n-x-1)$. Then 
	$
	{\textstyle\prod}_{\ell=x}^{n-1}(a+b+\ell)
	=
	{\textstyle\prod}_{\ell'=0}^{n-x-1}(a+b+x+\ell').
	$
	Therefore, the denominator of \eqref{eq:gamma_product_expansion} is 
	\begin{equation}
		{\textstyle\prod}_{m=0}^{n-1}(a+b+m)
		=
		\{{\textstyle\prod}_{m=0}^{x-1}(a+b+m)\}
		\{{\textstyle\prod}_{\ell'=0}^{n-x-1}(a+b+x+\ell')\}.
		\label{eq:denominator_factorization}
	\end{equation}
	
	From \eqref{eq:gamma_product_expansion} and \eqref{eq:denominator_factorization} we have
	\begin{equation}
		m_1^*(x\mid a,b)
		=
		\{{\textstyle\prod}_{j=0}^{x-1}
		\{(a+j)/(a+b+j)\}\}
		\{{\textstyle\prod}_{k=0}^{n-x-1}
		\{(b+k)/(a+b+x+k)\}\}.
		\label{eq:marginal_likelihood_product_form}
	\end{equation}
	
	Let
	$
	t_s=(a_s+b_s), p_s=a_s/(a_s+b_s),
	$
	and suppose that \(t_s\to\infty\) and \(p_s\to p\in[0,1]\). Then
	\(a_s=p_st_s\) and \(b_s=(1-p_s)t_s\).  
	Then for each fixed $j=0,\ldots,(x-1)$,
	$
	(a_s+j)/(a_s+b_s+j)
	=
	(p_st_s+j)/(t_s+j)
	\longrightarrow p,
	$
	and for each fixed $k=0,\ldots,(n-x-1)$, 
	$(b_s+k)/(a_s+b_s+x+k)
	=
	\{(1-p_s)t_s+k\}/(t_s+x+k)
	\longrightarrow 1-p.
	$
	
	\noindent Therefore
	\begin{equation}
		m_1^*(x\mid a_s,b_s)
		\longrightarrow
		p^x(1-p)^{n-x}
		=f(p).
		\label{eq:limit_of_marginal_likelihood}
	\end{equation}
	In particular, choosing \(p=p_0=x/n\) proves the sufficient part.
	We now prove the necessary part in the following.
	
	Suppose that a sequence $(a_s,b_s)\subset\mathcal H$ satisfies
	$
	m_1^*(x\mid a_s,b_s)\to M.
	$
	Assume to the contrary that $(a_s+b_s)$ does not diverge.  
	Then there exists a subsequence $(a_{s_k},b_{s_k})$ and a constant $L>0$ such that $(a_{s_k}+b_{s_k})\le L$.	
	If both $a_{s_k}$ and $b_{s_k}$ stay boundedly away from zero, then
	the points lie in a compact subset $K\subset\mathcal H$.
	Since Result 2.1 shows that
	$
	m_1^*(x\mid a,b)<M \text{ for every finite }(a,b)\in\mathcal H$, 
	continuity implies that $m_1^*$ attains its maximum on $K$ at some value
	$M_K<M$.
	Hence
	$
	m_1^*(x\mid a_{s_k},b_{s_k})
	\le M_K<M,
	$
	which contradicts the assumption that
	$m_1^*(x\mid a_s,b_s)\to M$.	
	Since $1\le x\le n-1$, the first product in
	\eqref{eq:marginal_likelihood_product_form}
	contains the factor corresponding to $j=0$, namely
	$
	a_{s_k}/(a_{s_k}+b_{s_k}).
	$
	If $a_{s_k}\to0$ while $b_{s_k}$ does not vanish, then
	$
	a_{s_k}/(a_{s_k}+b_{s_k})
	\longrightarrow 0.
	$
	Since all other factors in
	\eqref{eq:marginal_likelihood_product_form}
	are bounded above by $1$, it follows that
	$
	m_1^*(x\mid a_{s_k},b_{s_k})
	\longrightarrow 0,
	$
	contradicting $M>0$.
	
	If both $a_{s_k}\to0$ and $b_{s_k}\to0$, then
	the second product in
	\eqref{eq:marginal_likelihood_product_form}
	contains the factor corresponding to $k=0$,
	namely
	$
	b_{s_k}/(a_{s_k}+b_{s_k}+x),
	$
	which tends to $0$. Again all remaining factors are bounded above by $1$, so
	$
	m_1^*(x\mid a_{s_k},b_{s_k})
	\longrightarrow 0,
	$
	contradicting $M>0$.
	The case $b_{s_k}\to0$ is symmetric.  
	Hence $(a_s+b_s)\to\infty$.
	
	
	
	Let $t_s=(a_s+b_s), p_s=a_s/(a_s+b_s)$. 	
	Then $a_s=p_s t_s$ and $b_s=(1-p_s)t_s$, with $t_s\to\infty$.  
	If $p_s$ does not converge to $p_0=x/n$, then a subsequence $p_{s_k}\to p^*\neq p_0$ exists.
	Since $(a_s+b_s)\to\infty$ implies that
	$(a_{s_k}+b_{s_k})\to\infty$
	and
	$p_{s_k}\to p^*$,
	equation \eqref{eq:limit_of_marginal_likelihood} yields
	$
	m_1^*(x\mid a_{s_k},b_{s_k})
	\to
	(p^*)^x(1-p^*)^{n-x}
	$. 	
	But every subsequence must also converge to $M$, hence
	$
	(p^*)^x(1-p^*)^{n-x}=M.
	$
	Since the function $f(p)=p^x(1-p)^{n-x}$ attains its unique maximum at $p_0=x/n$, we obtain a contradiction.  
	Thus $a_s/(a_s+b_s)\to x/n$.
	
	Thus every sequence $(a_s,b_s)\subset\mathcal H$ satisfying
	$
	m_1^*(x\mid a_s,b_s)\to M
	$
	must satisfy
	$
	(a_s+b_s)\to\infty, a_s/(a_s+b_s)\to x/n.
	$	
	This completes the proof of Result 2.2.

	\subsection{Proof that $\ell'(c) > 0$ when $\Delta\ge0$}
	\label{app:delta_nonnegative}
	We consider two subcases.
	By symmetry, it suffices to consider $x\le n/2$.
	
	\paragraph{Subcase 1: $n=2m$.}
	
	The above restriction implies $x \le m = n/2$.
	For the third sum in \eqref{eq:lprime_c_harmonic}:
	$
	2\sum_{k=0}^{n-1} 1/(2c+k)
	=
	2\sum_{k=0}^{2m-1} 1/(2c+k).
	$
	Separating the even and odd terms yields
	\[
	2{\textstyle\sum}_{k=0}^{2m-1}\{1/(2c+k)\}
	=
	2{\textstyle\sum}_{r=0}^{m-1}\{1/(2c+2r)\}
	+
	2{\textstyle\sum}_{r=0}^{m-1}\{1/(2c+2r+1)\}.
	\]
	
	\noindent Hence
	\begin{equation}\label{eq:harmonic_even_odd_split}
		2{\textstyle\sum}_{k=0}^{2m-1}\{1/(2c+k)\}
		=
		{\textstyle\sum}_{r=0}^{m-1}\{1/(c+r)\}
		+
		{\textstyle\sum}_{r=0}^{m-1}\{1/(c+r+1/2)\}.
	\end{equation}
	
	For the second summation in \eqref{eq:lprime_c_harmonic}. Since $n=2m$, then
	$
	{\textstyle\sum}_{j=0}^{n-x-1}\{1/(c+j)\}
	=
	{\textstyle\sum}_{j=0}^{2m-x-1}\{1/(c+j)\}.
	$ Splitting the range at $m$ gives
	\begin{equation}\label{eq:harmonic_split_tail}
		{\textstyle\sum}_{j=0}^{2m-x-1}\{1/(c+j)\}
		=
		{\textstyle\sum}_{j=0}^{m-1}\{1/(c+j)\}
		+
		{\textstyle\sum}_{j=m}^{2m-x-1}\{1/(c+j)\}.
	\end{equation}
	
	\noindent Substituting \eqref{eq:harmonic_even_odd_split} and \eqref{eq:harmonic_split_tail} into the expression of $\ell'(c)$, we obtain
	\[
	\ell'(c)
	=
	\sum_{i=0}^{x-1}\frac{1}{c+i}
	+
	\sum_{j=0}^{m-1}\frac{1}{c+j}
	+
	\sum_{j=m}^{2m-x-1}\frac{1}{c+j}
	-
	\sum_{r=0}^{m-1}\frac{1}{c+r}
	-
	\sum_{r=0}^{m-1}\frac{1}{c+r+1/2}.
	\]
	
	\noindent After cancellation of the common summation terms, this simplifies to
	\begin{equation}\label{eq:lprime_m_reduction}
		\ell'(c)
		=
		{\textstyle\sum}_{i=0}^{x-1}\{1/(c+i)\}
		+
		{\textstyle\sum}_{j=m}^{2m-x-1}\{1/(c+j)\}
		-
		{\textstyle\sum}_{r=0}^{m-1}\{1/(c+r+1/2)\}.
	\end{equation}
	Define
	\[
	S=\{0,1,\ldots,(x-1),\, m, (m+1),\ldots,(2m-x-1)\},
	\]
	\[
	T=\{1/2,3/2,\ldots,(m-1/2)\}.
	\]
	
	\noindent Note that both $S$ and $T$ contain $m$ elements. Hence \eqref{eq:lprime_m_reduction} can be written as $
	\ell'(c)
	=
	{\textstyle\sum}_{s\in S}\{1/(c+s)\}
	-
	{\textstyle\sum}_{t\in T}\{1/(c+t)\}.
	$ Arrange the elements of $S$ and $T$ in increasing order: 
	$
	s_1<s_2<\cdots<s_m, t_1<t_2<\cdots<t_m
	$. Then
	\[
	s_i=
	\begin{cases}
		i-1, & 1\le i\le x,\\[4pt]
		m+i-x-1, & x<i\le m,
	\end{cases}
	\qquad
	t_i=i-1/2 .
	\]
	
	\noindent Define $
	D_k=\sum_{i=1}^{k} t_i-\sum_{i=1}^{k} s_i, k=1,2,\ldots,m .
	$ 
	We first show that $D_k\ge0$.
	
	\paragraph{If $1\le k\le x$,}
	
	in this range, $s_i=i-1, t_i=i-1/2$. Hence
	\begin{equation}\label{eq:Dk_positive}
		D_k
		=
		{\textstyle\sum}_{i=1}^{k}(i-1/2)
		-
		{\textstyle\sum}_{i=1}^{k}(i-1)
		=
		k/2>0.
	\end{equation}
	
	\paragraph{If $x<k\le m$,}
	
	then
	$
	\sum_{i=1}^{k} t_i
	=
	\sum_{i=1}^{k}\left(i-1/2\right)
	=
	k^2/2,
	$ and
	$
	\sum_{i=1}^{k}s_i
	=
	\sum_{i=1}^{x}(i-1)
	+
	\sum_{i=x+1}^{k}(m+i-x-1).
	$ 
	
	\noindent Note that
	$
	\sum_{i=x+1}^{k}(m+i-x-1)
	=
	(k-x)(m-x-1)+\sum_{i=x+1}^{k} i .
	$ Hence
	$
	\sum_{i=1}^{k}s_i
	=
	\sum_{i=1}^{x}(i-1)
	+
	(k-x)(m-x-1)
	+
	\sum_{i=x+1}^{k} i.
	$ After simplification,
	$
	\sum_{i=1}^{k}s_i
	=
	k^2/2
	+
	k(m-x-1/2)
	-
	x(m-x).
	$ Therefore
	\begin{equation}\label{eq:Dk_linear_form}
		D_k
		=
		{\textstyle\sum}_{i=1}^{k}t_i
		-
		{\textstyle\sum}_{i=1}^{k}s_i
		=
		k(x-m+1/2)
		+
		x(m-x).
	\end{equation}
	
	\noindent Since $x<k \le m$ and $x \in Z$, we have $	x-m+1/2 \le -1/2$, and therefore $D_k$ is a decreasing function of $k$ on $(x,m]$. Hence the minimum value of $D_k$ is attained at $k=m$. From \eqref{eq:Dk_linear_form} we have
	$
	D_m
	=
	m(x-m+1/2)+x(m-x).
	$
	A direct simplification yields
	$
	D_m
	=
	n/4-(x-n/2)^2
	=
	\Delta .
	$
	Since we are considering the case $\Delta\ge0$, it follows that $D_m\ge0.$
	Thus $D_k$ is decreasing in $k$ on $x<k\le m$. Hence its minimum over this range is attained at $k=m$. Since $D_m=\Delta\ge0$, it follows that $D_k\ge0$ for all $k=x+1,\ldots,m$. Together with the result for $1\le k\le x$ obtained in \eqref{eq:Dk_positive}, we conclude that
	\begin{equation}\label{eq:Dk_positive_all}
		D_k\ge0,
		\qquad
		k=1,2,\ldots,m.
	\end{equation}
	
	Let
	$
	f(u)=1/(c+u), u>-c
	$. Then
	$
	f'(u)=-1/(c+u)^2<0, f''(u)=2/(c+u)^3>0.
	$ Hence $f(u)$ is strictly convex and strictly decreasing. From \eqref{eq:Dk_positive_all}, we have
	$
	\sum_{i=1}^k t_i\ge \sum_{i=1}^k s_i,\qquad k=1,\ldots,m.
	$
	Thus the ordered sequence $\{t_i\}_{i=1}^m$ weakly majorizes $\{s_i\}_{i=1}^m$ in the sense of partial sums.
	Since $f(u)=1/(c+u)$ is strictly convex and strictly decreasing, the weak Karamata inequality gives
	$
	\sum_{i=1}^{m} f(s_i)\ge \sum_{i=1}^{m} f(t_i).
	$
	Moreover, the two ordered sequences are not identical, so the inequality is strict:
	$
	\sum_{i=1}^{m} f(s_i)> \sum_{i=1}^{m} f(t_i)
	$. 
	Substituting $f(u)=1/(c+u)$ gives $\sum_{i=1}^{m}1/(c+s_i)
	>
	\sum_{i=1}^{m}1/(c+t_i)$. Hence
	$
	\ell'(c)
	=
	\sum_{i=1}^{m} 1/(c+s_i)
	-
	\sum_{i=1}^{m} 1/(c+t_i)
	>0.
	$
	
	\paragraph{Subcase 2: $n=2m+1$.}
	
	Due to the symmetry $x \leftrightarrow n-x$, it suffices
	to consider the case $x \le n/2$. Since $n=2m+1$ and $x\in\mathbb{Z}_+$, this implies $x \le m$. From \eqref{eq:lprime_c_harmonic} and $n=2m+1$, we obtain
	\begin{equation}\label{eq:lprime_odd_n}
		\ell'(c)
		=
		\sum_{i=0}^{x-1}\frac{1}{c+i}
		+
		\sum_{j=0}^{2m-x}\frac{1}{c+j}
		-
		2\sum_{k=0}^{2m}\frac{1}{2c+k}.
	\end{equation}
	
	\noindent Consider the third term on the RHS of \eqref{eq:lprime_odd_n}. Splitting even and odd indices gives
	\[
	2{\textstyle\sum}_{k=0}^{2m}\{1/(2c+k)\}
	=
	2{\textstyle\sum}_{r=0}^{m}\{1/(2c+2r)\}
	+
	2{\textstyle\sum}_{r=0}^{m-1}\{1/(2c+2r+1)\}.
	\]
	
	\noindent Hence
	\[
	2{\textstyle\sum}_{k=0}^{2m}\{1/(2c+k)\}
	=
	{\textstyle\sum}_{r=0}^{m}\{1/(c+r)\}
	+
	{\textstyle\sum}_{r=0}^{m-1}\{1/(c+r+1/2)\}.
	\]
	\noindent Next consider
	$
	{\textstyle\sum}_{j=0}^{2m-x}\{1/(c+j)\}
	=
	{\textstyle\sum}_{j=0}^{m}\{1/(c+j)\}
	+
	{\textstyle\sum}_{j=m+1}^{2m-x}\{1/(c+j)\}.
	$
	Substituting these decompositions into \eqref{eq:lprime_odd_n} yields
	\begin{equation}\label{eq:lprime_odd_reduction}
		\ell'(c)
		=
		{\textstyle\sum}_{i=0}^{x-1}\{1/(c+i)\}
		+
		{\textstyle\sum}_{j=m+1}^{2m-x}\{1/(c+j)\}
		-
		{\textstyle\sum}_{r=0}^{m-1}\{1/(c+r+1/2)\}.
	\end{equation}
	
	\noindent Define the sets
	\[
	S=\{0,1,\ldots,(x-1),\,(m+1),(m+2),\ldots,(2m-x)\},
	\]
	\[
	T=\{1/2,3/2,\ldots,m-(1/2)\}.
	\]
	
	\noindent Both $S$ and $T$ contain $m$ elements, and \eqref{eq:lprime_odd_reduction} can be written as
	$
	\ell'(c)
	=
	{\textstyle\sum}_{s\in S}\{1/(c+s)\}
	-
	{\textstyle\sum}_{t\in T}\{1/(c+t)\}.
	$ 
	Write the elements of $S$ and $T$ in increasing order:
	\[
	s_i=
	\begin{cases}
		i-1, & 1\le i\le x,\\[4pt]
		m+i-x, & x<i\le m,
	\end{cases}
	\qquad
	t_i=i-1/2.
	\]
	Define
	\[
	D_k=\sum_{i=1}^{k}t_i-\sum_{i=1}^{k}s_i,
	\qquad k=1,\ldots,m.
	\]
	If $1\le k\le x$, then $D_k=k/2>0$. If $x<k\le m$, then
	$
	D_k=k(x-m-1/2)+x(m-x+1).
	$
	Thus $D_k$ is decreasing in $k$ on $x<k\le m$, so its minimum is attained at $k=m$. Moreover,
	$
	D_m
	=
	m(x-m-1/2)+x(m-x+1)
	=
	n/4-(x-n/2)^2
	=
	\Delta.
	$
	Since $\Delta\ge0$, it follows that
	$
	D_k\ge0, k=1,\ldots,m.
	$
	By the weak Karamata inequality of Marshall et al. (2011), we obtain
	$
	\ell'(c)
	=
	{\textstyle\sum}_{s\in S}\{1/(c+s)\}
	-
	{\textstyle\sum}_{t\in T}\{1/(c+t)\}
	>0, \text{for all } c>0.
	$
	Hence $\ell(c)$ is strictly increasing on $(0,\infty)$, and
	$m_1^*(x\mid c,c)$ has no interior maximizer.

	
	\newpage
	\subsection{Numerically Determined Maximizers}
	\label{app:numerical_maximizers}
	
	Table~\ref{tab:c_hat_theta_hat_appendix}
	reports the numerically determined maximizing values of $c$
	and the corresponding empirical Bayes estimates
	for several values of $n$.
	Only the cases
	$0\leq x\leq\lfloor n/2\rfloor$
	are displayed because the remaining cases follow by symmetry.
	
	\begin{table}[H]
		\centering
		
		\rotatebox{90}{%
			\begin{minipage}{0.88\textheight}
				\centering
				
				\setlength{\abovecaptionskip}{0pt}
				\setlength{\belowcaptionskip}{3pt}
				
				\caption{Numerically determined values of $\widehat{c_n}$ and the
					corresponding empirical Bayes estimator $\widehat{\theta}_{EB}$.
					Only $0\leq x\leq\lfloor n/2\rfloor$ are displayed because
					the remaining values follow by symmetry.}
				\label{tab:c_hat_theta_hat_appendix}
				
				\footnotesize
				\setlength{\tabcolsep}{2.8pt}
				\renewcommand{\arraystretch}{0.80}
				
				\begin{adjustbox}{
						max width=\linewidth,
						max totalheight=0.82\textwidth,
						center
					}
					
					\begin{tabular}{c|*{10}{cc}}
						\hline
						
						\multirow{2}{*}{$x\backslash n$}
						& \multicolumn{2}{c}{$2$}
						& \multicolumn{2}{c}{$5$}
						& \multicolumn{2}{c}{$10$}
						& \multicolumn{2}{c}{$20$}
						& \multicolumn{2}{c}{$30$}
						& \multicolumn{2}{c}{$40$}
						& \multicolumn{2}{c}{$50$}
						& \multicolumn{2}{c}{$60$}
						& \multicolumn{2}{c}{$70$}
						& \multicolumn{2}{c}{$80$}
						\\
						\cline{2-21}
					
					& $\widehat c$ & $\widehat\theta$
					& $\widehat c$ & $\widehat\theta$
					& $\widehat c$ & $\widehat\theta$
					& $\widehat c$ & $\widehat\theta$
					& $\widehat c$ & $\widehat\theta$
					& $\widehat c$ & $\widehat\theta$
					& $\widehat c$ & $\widehat\theta$
					& $\widehat c$ & $\widehat\theta$
					& $\widehat c$ & $\widehat\theta$
					& $\widehat c$ & $\widehat\theta$
					\\
					\hline
					
					$0$
					& $0^+$ & $0$
					& $0^+$ & $0$
					& $0^+$ & $0$
					& $0^+$ & $0$
					& $0^+$ & $0$
					& $0^+$ & $0$
					& $0^+$ & $0$
					& $0^+$ & $0$
					& $0^+$ & $0$
					& $0^+$ & $0$
					\\
					
					$1$
					& $\infty$ & $0.5$
					& $2.427$ & $0.348$
					& $0.670$ & $0.147$
					& $0.411$ & $0.068$
					& $0.340$ & $0.044$
					& $0.303$ & $0.032$
					& $0.281$ & $0.025$
					& $0.265$ & $0.021$
					& $0.253$ & $0.018$
					& $0.243$ & $0.015$
					\\
					
					$2$
					& {} & {}
					& $\infty$ & $0.5$
					& $1.648$ & $0.274$
					& $0.653$ & $0.125$
					& $0.491$ & $0.080$
					& $0.419$ & $0.059$
					& $0.377$ & $0.047$
					& $0.349$ & $0.039$
					& $0.328$ & $0.033$
					& $0.312$ & $0.029$
					\\
					
					$3$
					& {} & {}
					& {} & {}
					& $7.475$ & $0.420$
					& $0.969$ & $0.181$
					& $0.650$ & $0.117$
					& $0.530$ & $0.086$
					& $0.464$ & $0.068$
					& $0.422$ & $0.056$
					& $0.392$ & $0.048$
					& $0.370$ & $0.042$
					\\
					
					$4$
					& {} & {}
					& {} & {}
					& $\infty$ & $0.5$
					& $1.454$ & $0.238$
					& $0.839$ & $0.153$
					& $0.649$ & $0.113$
					& $0.553$ & $0.089$
					& $0.494$ & $0.074$
					& $0.454$ & $0.063$
					& $0.424$ & $0.055$
					\\
					
					$5$
					& {} & {}
					& {} & {}
					& $\infty$ & $0.5$
					& $2.322$ & $0.297$
					& $1.080$ & $0.189$
					& $0.784$ & $0.139$
					& $0.648$ & $0.110$
					& $0.569$ & $0.091$
					& $0.515$ & $0.078$
					& $0.477$ & $0.068$
					\\
					
					$6$
					& {} & {}
					& {} & {}
					& {} & {}
					& $4.285$ & $0.360$
					& $1.404$ & $0.226$
					& $0.944$ & $0.166$
					& $0.754$ & $0.131$
					& $0.648$ & $0.108$
					& $0.580$ & $0.092$
					& $0.532$ & $0.081$
					\\
					
					$7$
					& {} & {}
					& {} & {}
					& {} & {}
					& $11.860$ & $0.431$
					& $1.866$ & $0.263$
					& $1.138$ & $0.192$
					& $0.873$ & $0.152$
					& $0.734$ & $0.126$
					& $0.648$ & $0.107$
					& $0.588$ & $0.093$
					\\
					
					$8$
					& {} & {}
					& {} & {}
					& {} & {}
					& $\infty$ & $0.5$
					& $2.576$ & $0.301$
					& $1.381$ & $0.219$
					& $1.011$ & $0.173$
					& $0.830$ & $0.143$
					& $0.721$ & $0.122$
					& $0.648$ & $0.106$
					\\
					
					$9$
					& {} & {}
					& {} & {}
					& {} & {}
					& $\infty$ & $0.5$
					& $3.783$ & $0.340$
					& $1.695$ & $0.246$
					& $1.173$ & $0.194$
					& $0.936$ & $0.161$
					& $0.800$ & $0.137$
					& $0.711$ & $0.119$
					\\
					
					$10$
					& {} & {}
					& {} & {}
					& {} & {}
					& $\infty$ & $0.5$
					& $6.192$ & $0.382$
					& $2.116$ & $0.274$
					& $1.368$ & $0.216$
					& $1.058$ & $0.178$
					& $0.887$ & $0.152$
					& $0.779$ & $0.132$
					\\
					
					$11$
					& {} & {}
					& {} & {}
					& {} & {}
					& {} & {}
					& $12.781$ & $0.428$
					& $2.706$ & $0.302$
					& $1.606$ & $0.237$
					& $1.197$ & $0.195$
					& $0.984$ & $0.167$
					& $0.853$ & $0.145$
					\\
					
					$12$
					& {} & {}
					& {} & {}
					& {} & {}
					& {} & {}
					& $72.497$ & $0.483$
					& $3.579$ & $0.330$
					& $1.903$ & $0.258$
					& $1.359$ & $0.213$
					& $1.092$ & $0.181$
					& $0.933$ & $0.158$
					\\
					
					$13$
					& {} & {}
					& {} & {}
					& {} & {}
					& {} & {}
					& $\infty$ & $0.5$
					& $4.976$ & $0.360$
					& $2.283$ & $0.280$
					& $1.551$ & $0.231$
					& $1.214$ & $0.196$
					& $1.021$ & $0.171$
					\\
					
					$14$
					& {} & {}
					& {} & {}
					& {} & {}
					& {} & {}
					& $\infty$ & $0.5$
					& $7.482$ & $0.391$
					& $2.784$ & $0.302$
					& $1.780$ & $0.248$
					& $1.353$ & $0.211$
					& $1.119$ & $0.184$
					\\
					
					$15$
					& {} & {}
					& {} & {}
					& {} & {}
					& {} & {}
					& $\infty$ & $0.5$
					& $12.988$ & $0.424$
					& $3.468$ & $0.324$
					& $2.058$ & $0.266$
					& $1.513$ & $0.226$
					& $1.227$ & $0.197$
					\\
					
					$16$
					& {} & {}
					& {} & {}
					& {} & {}
					& {} & {}
					& {} & {}
					& $32.494$ & $0.462$
					& $4.445$ & $0.347$
					& $2.402$ & $0.284$
					& $1.699$ & $0.241$
					& $1.349$ & $0.210$
					\\
					
					$17$
					& {} & {}
					& {} & {}
					& {} & {}
					& {} & {}
					& {} & {}
					& $\infty$ & $0.5$
					& $5.926$ & $0.371$
					& $2.837$ & $0.302$
					& $1.918$ & $0.256$
					& $1.487$ & $0.223$
					\\
					
					$18$
					& {} & {}
					& {} & {}
					& {} & {}
					& {} & {}
					& {} & {}
					& $\infty$ & $0.5$
					& $8.375$ & $0.395$
					& $3.399$ & $0.320$
					& $2.178$ & $0.271$
					& $1.643$ & $0.236$
					\\
					
					$19$
					& {} & {}
					& {} & {}
					& {} & {}
					& {} & {}
					& {} & {}
					& $\infty$ & $0.5$
					& $13.021$ & $0.421$
					& $4.148$ & $0.339$
					& $2.491$ & $0.287$
					& $1.823$ & $0.249$
					\\
					
					$20$
					& {} & {}
					& {} & {}
					& {} & {}
					& {} & {}
					& {} & {}
					& $\infty$ & $0.5$
					& $24.493$ & $0.449$
					& $5.184$ & $0.358$
					& $2.874$ & $0.302$
					& $2.030$ & $0.262$
					\\
					
					$21$
					& {} & {}
					& {} & {}
					& {} & {}
					& {} & {}
					& {} & {}
					& {} & {}
					& $87.498$ & $0.482$
					& $6.687$ & $0.377$
					& $3.351$ & $0.317$
					& $2.273$ & $0.275$
					\\
					
					$22$
					& {} & {}
					& {} & {}
					& {} & {}
					& {} & {}
					& {} & {}
					& {} & {}
					& $\infty$ & $0.5$
					& $9.017$ & $0.397$
					& $3.958$ & $0.333$
					& $2.560$ & $0.289$
					\\
					
					$23$
					& {} & {}
					& {} & {}
					& {} & {}
					& {} & {}
					& {} & {}
					& {} & {}
					& $\infty$ & $0.5$
					& $13.004$ & $0.419$
					& $4.750$ & $0.349$
					& $2.902$ & $0.302$
					\\
					
					$24$
					& {} & {}
					& {} & {}
					& {} & {}
					& {} & {}
					& {} & {}
					& {} & {}
					& $\infty$ & $0.5$
					& $21.064$ & $0.441$
					& $5.814$ & $0.365$
					& $3.317$ & $0.315$
					\\
					
					$25$
					& {} & {}
					& {} & {}
					& {} & {}
					& {} & {}
					& {} & {}
					& {} & {}
					& $\infty$ & $0.5$
					& $44.246$ & $0.466$
					& $7.302$ & $0.382$
					& $3.827$ & $0.329$
					\\
					
					$26$
					& {} & {}
					& {} & {}
					& {} & {}
					& {} & {}
					& {} & {}
					& {} & {}
					& {} & {}
					& $442.499$ & $0.496$
					& $9.495$ & $0.399$
					& $4.465$ & $0.343$
					\\
					
					$27$
					& {} & {}
					& {} & {}
					& {} & {}
					& {} & {}
					& {} & {}
					& {} & {}
					& {} & {}
					& $\infty$ & $0.5$
					& $12.974$ & $0.417$
					& $5.282$ & $0.356$
					\\
					
					$28$
					& {} & {}
					& {} & {}
					& {} & {}
					& {} & {}
					& {} & {}
					& {} & {}
					& {} & {}
					& $\infty$ & $0.5$
					& $19.159$ & $0.435$
					& $6.354$ & $0.371$
					\\
					
					$29$
					& {} & {}
					& {} & {}
					& {} & {}
					& {} & {}
					& {} & {}
					& {} & {}
					& {} & {}
					& $\infty$ & $0.5$
					& $32.630$ & $0.456$
					& $7.807$ & $0.385$
					\\
					
					$30$
					& {} & {}
					& {} & {}
					& {} & {}
					& {} & {}
					& {} & {}
					& {} & {}
					& {} & {}
					& $\infty$ & $0.5$
					& $80.498$ & $0.478$
					& $9.863$ & $0.400$
					\\
					
					$31$
					& {} & {}
					& {} & {}
					& {} & {}
					& {} & {}
					& {} & {}
					& {} & {}
					& {} & {}
					& {} & {}
					& $\infty$ & $0.5$
					& $12.941$ & $0.415$
					\\
					
					$32$
					& {} & {}
					& {} & {}
					& {} & {}
					& {} & {}
					& {} & {}
					& {} & {}
					& {} & {}
					& {} & {}
					& $\infty$ & $0.5$
					& $17.947$ & $0.431$
					\\
					
					$33$
					& {} & {}
					& {} & {}
					& {} & {}
					& {} & {}
					& {} & {}
					& {} & {}
					& {} & {}
					& {} & {}
					& $\infty$ & $0.5$
					& $27.236$ & $0.448$
					\\
					
					$34$
					& {} & {}
					& {} & {}
					& {} & {}
					& {} & {}
					& {} & {}
					& {} & {}
					& {} & {}
					& {} & {}
					& $\infty$ & $0.5$
					& $49.371$ & $0.466$
					\\
					
					$35$
					& {} & {}
					& {} & {}
					& {} & {}
					& {} & {}
					& {} & {}
					& {} & {}
					& {} & {}
					& {} & {}
					& $\infty$ & $0.5$
					& $157.999$ & $0.487$
					\\
					
					$36$
					& {} & {}
					& {} & {}
					& {} & {}
					& {} & {}
					& {} & {}
					& {} & {}
					& {} & {}
					& {} & {}
					& {} & {}
					& $\infty$ & $0.5$
					\\
					
					$37$
					& {} & {}
					& {} & {}
					& {} & {}
					& {} & {}
					& {} & {}
					& {} & {}
					& {} & {}
					& {} & {}
					& {} & {}
					& $\infty$ & $0.5$
					\\
					
					$38$
					& {} & {}
					& {} & {}
					& {} & {}
					& {} & {}
					& {} & {}
					& {} & {}
					& {} & {}
					& {} & {}
					& {} & {}
					& $\infty$ & $0.5$
					\\
					
					$39$
					& {} & {}
					& {} & {}
					& {} & {}
					& {} & {}
					& {} & {}
					& {} & {}
					& {} & {}
					& {} & {}
					& {} & {}
					& $\infty$ & $0.5$
					\\
					
					$40$
					& {} & {}
					& {} & {}
					& {} & {}
					& {} & {}
					& {} & {}
					& {} & {}
					& {} & {}
					& {} & {}
					& {} & {}
					& $\infty$ & $0.5$
					\\
					
					\hline
				\end{tabular}
			\end{adjustbox}
		\end{minipage}%
	}
\end{table}

\section*{References}

	\begin{enumerate}
		
		\item Abramowitz, M. and Stegun, I. A. (Eds.) (1964).
		\textit{Handbook of Mathematical Functions with Formulas, Graphs, and Mathematical Tables}.
		National Bureau of Standards, Washington, DC, USA.
		
		\item Arellano-Valle, R. B., Ferreira, C. S., and Genton, M. G. (2018).
		Scale and shape mixtures of multivariate skew-normal distributions.
		\textit{Journal of Multivariate Analysis},
		166, 98--110.

		\item Azzalini, A. (2013).
		\textit{The Skew-Normal and Related Families}.
		In collaboration with Antonella Capitanio.
		Cambridge University Press.
		\url{https://doi.org/10.1017/CBO9781139248891}.

		\item Berger, J. O. (1980).
		Improving on inadmissible estimators in continuous exponential families
		with applications to simultaneous estimation of Gamma scale parameters.
		\textit{The Annals of Statistics}, 8, 545--571.

		\item Berger, J. O. (1985).
		\textit{Statistical Decision Theory and Bayesian Analysis} (2nd ed.).
		Springer-Verlag, New York, USA.

		\item Carlin, B. P. and Louis, T. A. (2000).
		Empirical Bayes: Past, present and future.
		\textit{Journal of the American Statistical Association},
		95(452), 1286--1289.

		\item Casella, G. (1985).
		An introduction to empirical Bayes data analysis.
		\textit{The American Statistician}, 39(2), 83--87.

		\item Chen, Y. and Lei, L. (2025).
		Compound estimation for binomials.
		\textit{arXiv preprint arXiv:2512.25042}.

		\item Doss, H. and Linero, A. R. (2024).
		Scalable empirical Bayes inference and Bayesian sensitivity analysis.
		\textit{Statistical Science}, 39(4), 601--622.

		\item Efron, B. (2019).
		Bayes, Oracle Bayes, and empirical Bayes.
		\textit{Statistical Science}, 34(2), 177--201.

		\item Efron, B. and Morris, C. (1975).
		Data analysis using Stein's estimator and its generalizations.
		\textit{Journal of the American Statistical Association},
		70(350), 311--319.

		\item Gelman, A., Carlin, J. B., Stern, H. S., Dunson, D. B.,
		Vehtari, A., and Rubin, D. B. (2013).
		\textit{Bayesian Data Analysis} (3rd ed.).
		CRC Press, Boca Raton, FL, USA.

		\item Ghosh, M. and Parsian, A. (1981).
		Bayes minimax estimation of multiple Poisson parameters.
		\textit{Journal of Multivariate Analysis}, 11, 280--288.

		\item Gutmann, S. (1982).
		Minimax linear empirical Bayes estimation of the binomial parameter.
		\textit{Communications in Statistics -- Theory and Methods},
		11(18), 2075--2082.

		\item Jana, S., Polyanskiy, Y., and Wu, Y. (2025).
		Optimal empirical Bayes estimation for the Poisson model via
		minimum-distance methods.
		\textit{Information and Inference: A Journal of the IMA},
		14(4), iaaf027.

		\item Johnson, B. McK. (1971).
		On the admissible estimators for certain fixed sample binomial problems.
		\textit{The Annals of Mathematical Statistics},
		42(5), 1579--1587.

		\item Kang, B., Polyanskiy, Y., and Teh, A. (2026).
		Function estimation in the empirical Bayes setting.
		\textit{arXiv preprint arXiv:2601.18689}.

		\item Maritz, J. S. and Lwin, T. (1989).
		\textit{Empirical Bayes Methods} (2nd ed.).
		CRC Press, Boca Raton, Florida, USA.

		\item Marshall, A. W., Olkin, I., and Arnold, B. C. (2011).
		\textit{Inequalities: Theory of Majorization and Its Applications}
		(2nd ed.).
		Springer, New York, USA.

		\item Martz, H. F. and Lian, M. G. (1974).
		Empirical Bayes estimation of the binomial parameter.
		\textit{Biometrika}, 61(3), 517--523.

		\item Robbins, H. (1956).
		An empirical Bayes approach to statistics.
		In J. Neyman (Ed.),
		\textit{Proceedings of the Third Berkeley Symposium on Mathematical
			Statistics and Probability, Volume I: Contributions to the Theory of
			Statistics}, 157--163.
		University of California Press, Berkeley, USA.
		
	\end{enumerate}
\end{document}